\documentstyle[prb,aps,array,epsfig]{revtex}
\begin{document}

\title{The microscopic theory of superfluid $^4$He}


\author{J. X. Zheng-Johansson$^1$\footnote{Email: jxzj@mailaps.org}, B. Johansson$^2$, and P-I. Johansson$^3$} 
\address{1. H. H. Wills Physics Laboratory, Bristol University, Tyndall Avenue, Bristol, BS8 1TL, England \\
2. Condensed Matter Theory Group, Department of Physics, Uppsala  University, Box 530, 75121 Uppsala, Sweden\\
3. Department of Neutron Research, Uppsala University, Studsvik 611 82, Sweden}
 \date{October, 2002}

\def\Asw{A_{{\rm sw}}}
\def\g{\gamma}
\def\OAM{O_{_{AM}}}
\def\OUM{O_{_{UM}}}
\def\OSM{O_{_{SM}}}
\def\OCM{O_{_{CM}}}

\def\xT1{T_{\lambda}^{(1)}} 
\def\Tl2{T_{\lambda}^{(2)}} 
\def\a{\alpha}
\def\aon{\alpha_1}
\def\W{\Omega}
\def\Us0{U_{s0}}
\def\Un0{U_{n0}}
\def\Up{\Upsilon}
\def\Ph0{\Phi_{0}}
\def\Pha0{\Phi_{\alpha 0}}
\def\Vs0{V_{s0}}
\def\Db{\Delta_b}
\def\Ds{\Delta_s}
\def\Do{\Delta_0}
\def\Dsa{\Delta_{s\alpha}}
\def\Dsn{\Delta_{sn}}
\def\Dt{\widetilde{\Delta}}
\def\D{\Delta}
\def\Dto{\widetilde{\Delta}_0}
\def\eR{\subset}
\def\etae{\eta_{ev}}
\def\eRo{\subset_0}
\def\dn{\delta_{{\rm vn}}}
\def\dv{\delta_{{\rm v}}}
\def\ds{\delta_{s}}
\def\da{\delta_{\alpha}}
\def\us{\overline{u}_s }
\def\un{\overline{u}_n }
\def\ua{\overline{u}_{\alpha} }
\def\4He{$^4$He}
\def\la{\lambda}
\def\AAi{\AA$^{-1}$}
\def\uls{u_{l_{\Sigma}}}
\def\uons{u_{1_{\Sigma}}}
\def\u2s{u_{2_{\Sigma}}}
\def\uNs{u_{N_{\Upsilon, \Sigma}}}
\def\ulsp{u_{l'_{\Sigma}}}
\def\alt{{{}\atop{{<\atop\sim}\atop{}}} }
\def\len{7.5 cm}  
\def\lat{{\rm lat}}
\def\La{\Lambda}
\def\Bbf{${\bf B}$}
\def\Bbfa{${\bf B}_a$}
\def\Bbfind{${\bf B}_{ind}$}
\def\Ebf{${\bf E}$}
\def\BbfLd{${\bf B_{Ld}}$}
\def\Ebfa{${\bf E}_a$}
\def\Ebfind{${\bf E}_{ind}$}
\def\FS{F_{\Sigma}}
\def\HS{H_{\Sigma}}
\def\VS{V_{\Sigma}}
\def\Jbf{${\bf J}$}
\def\Jbfa{${\bf J}_a$}
\def\Jbfind{${\bf J}_{ind}$}
\def\Jbf{${\bf J}$}
\def\Jbfs{${\bf J_s}$}
\def\kbf{{\bf $ k $}}
\def\kbfsu{{\bf $ k \uparrow $}} 
\def\kbfsd{{\bf $ k \downarrow $}}
\def\kne{k_{{\rm ne}}} 
\def\oK{$^\circ$K} 
\def\cry{{\rm cry}}
\def\lf{\left}
\def\rt{\right}
\def\rbf{{\bf $ r $}}
\def\krot{\kappa_{{\rm rot}}}
\def\xbf{{\bf $ x $}}
\def\e{\epsilon}
\def\ev{\varepsilon}
\def\elg{{\LARGE e}}
\def\glg{{\LARGE g}}
\def\nlg{{\LARGE n}}
\def\slg{{\LARGE s}}
\def\sg{\sigma}
\def\b[{  {\huge [}  }
\def\Bb{{\bf B}}
\def\Bba{{\bf B}_a}
\def\Bbind{{\bf B}_{ind}}
\def\BbLd{{\bf B_{Ld}}}
\def\E{{\cal E}}
\def\Exm{\widetilde{{\cal E}}_x}
\def\Eym{\widetilde{{\cal E}}_y}
\def\Ezm{\widetilde{{\cal E}}_z}
\def\EK{{\cal E}(K)}
\def\Eq{{\cal E}(q)}
\def\Eb{{\bf E}}
\def\Eba{{\bf E}_a}
\def\Ebind{{\bf E}_{ind}}
\def\Jb{{\bf J}}
\def\Jba{{\bf J}_a}   
\def\Jbind{{\bf  J}_{ind}}
\def\Jb{{\bf J}}
\def\Jbs{{\bf J_s}}
\def\jbs{{\bf j_s}}
\def\jb{{\bf  j}}
\def\qb{{\bf  q}}
\def\bb{{\bf  b}}
\def\Gb{{\bf  G}}
\def\He4{$^4$He}
\def\Kb{{\bf  K}}
\def\An{$\mathop{\rm A}\limits^\circ$}
\def\Ans{\mathop{\rm A}\limits^\circ}
\def\Tc{ $T_c$}
\def\T1{ $T$}
\def\Tlmb{$T_\lambda$}
\def\Th{\mit \Theta}
\def\VRo{\forall_0}
\def\kb{{\bf  k }}
\def\kbsu{{\bf  k \uparrow }}
\def\kbsd{{\bf  k \downarrow }}
\def\lx{l_x}
\def\ly{l_y}
\def\lz{l_z}
\def\Lov{\overline{L}}
\def\Lw{L_{{\rm w}}}
\def\nnr{\nonumber}
\def\Nc{{\cal N}}
\def\Nu{N_{\Upsilon}}
\def\neu{{\rm ne}}

\def\Nlmb{N_{\lambda}}
\def\olZK{\overline  {Z(K)} }  
\def\ov{\over}
\def\pd{\partial}
\def\ph{{\rm ph}}
\def\rb{{\bf  r }}
\def\Rb{{\bf  R }}
\def\ub{{\bf  u }}
\def\vb{{\bf  v }}
\def\Vp{V}

\def\Vx{V_x}
\def\V{\cal {V}}
\def\tauw{\tau_{{\rm w}}}
\def\Tlm{T_{\lambda}}
\def\w{\omega}
\def\wl{\omega_l}
\def\wxm{\widetilde{\omega}_x}
\def\wym{\widetilde{\omega}_y}
\def\wzm{\widetilde{\omega}_z}
\def\xb{{\bf  x }}
\def\xib{\mbox{\boldmath$\xi$}}
\def\Xib{{\bf  \Xi }}
\def\itS{{\it  S }}
\def\itW{{\it  W }} \def\divd{\bigtriangledown } 
\def\divu{\bigtriangleup} 
\def\divdb{{\bf \bigtriangledown }} \def\divub{{\bf \bigtriangleup }} 
\def\calZ{{\cal Z}}

\def\Kcal{ \widetilde{K}}
\def\Kl{K_l}
\def\Kla{K_{l_\alpha}}
\def\Klx{K_{l_x}}
\def\Kly{K_{l_y}}
\def\Klz{K_{l_z}}
\def\Ka{K_{\alpha}}

\def\Kxh{ \widehat{K}_{x}}
\def\Kyh{ \widehat{K}_{y}}
\def\Kzh{ \widehat{K}_{z}}
\def\1N1D{{\cal N}_{1}}
\def\N2{{\cal N}_{2}}
\def\uncN{{\cal N}_{uc}}
\def\uN{{\cal N}_{u}}

\def\D2{D_2}
\def\xD1{D_1}
\def\uD{D_u}
\def\D{{\cal D}}
\def\Dc{{\cal D}}
\def\Dco{{\cal D}_0}

\def\Kam{ {\widetilde{K}}_\alpha}
\def\Kxm{ {\widetilde{K}}_x}
\def\Kym{ {\widetilde{K}}_y}
\def\Kzm{ {\widetilde{K}}_z}

\maketitle  

\begin{abstract}
We present a microscopic theory of superfluid $^4$He, formulated using the overall experimental observations as input information. With the theory of a consistent basis, we answer all of the essential questions regarding He II. To conform to the overall key experimental observations, we show that in their coexistence at a higher $T$ the superfluid and normal fluid atoms necessarily aggregate in separate large regions, which with falling $T$ reduces to predominantly superfluid aggregates. In a superfluid aggregate, termed a {\bf momenton}, the helium atomic wave-packets primarily perform {\bf localized (harmonic) oscillations} about sites fixed on the momenton, facilitating excitation states of discrete energy levels. The predominant thermal excitation in the superfluid equilibrium is hence between these discrete levels, i.e. in the form of single phonons. Superposed to the coherent oscillations of the single atoms above, the {\bf momenton waves}, comprising the {\it in phase, simultaneous} oscillations of all of the atoms in each momenton, give rise to the so-called second sound wave, which contribution to the total excitation and energy is  however  negligible. The superfluid atoms are strongly correlated, due to a many-quantum-atom correlation, having a binding energy $-$7.2 \oK/atom as from experiments; from this we suitably define a {\bf superfluid bond}, $\Delta_b\simeq $7.2 \oK/atom. $\Delta_b$ is substantial, on the scale of total excitation energy ($\sim$ 15 \oK) \ or of the binding energy of He I ($\sim 3$ \oK/atom), and inevitably dominates the phase stability of He II. Accordingly in the superfluid in {\it quasi equilibrium} between 0.6-2.17 \oK, thermal excitation is predominantly due to the break-up of $\Delta_b$, at an activation energy $\Delta_b + \dv \simeq 8.6$ \oK. This yields the experimentally observed $\lambda$ specific heat,  and in turn the {\bf inelastic damped-Bragg scattering} of neutrons at $1.93$ \AA$^{-1}$.  By embodying these concepts pertaining to static structure of the fluid, atomic bobnding and atomic dynamics, we establish equations of motion for the superfluid atoms, both  semi-classical and quantum mechanical, in a coordinate decomposition scheme. Based on their solutions, and the results from a corresponding statistical thermodynamic treatment, we have been able to quantitatively predict all the essential properties of He II, including the full excitation spectrum (dividing in $q< 1.93-\sg/2$, $1.93-\sg/2<1.93$ and $q=1.93$ \AAi), the two fluid fractions,  the specific heat (dividing in  the $T^3$- and the $\lambda$-$C_V$ regions), the superfluid phase transition temperature \Tlmb,  \ and in a separate paper II the critical velocity $v_c$, the total number of excitation states and the superfluid viscosity; all of which are in good overall good agreements with experiments. We have also evaluated the zero point energy, the potential energy, the first order thermodynamic functions including $U(T)$, $A(T)$, $S(T)$ and $J(T)$ of which $S(T)$ can be satisfactorily compared with experimental data; we also formally present the continuity and stability conditions of the superfluid, and the dynamics structure factor due to the $\Db$ excitation. The microscopic scheme  also facilitates a novel, consistent {\bf QCE superfluidity mechanism}, presented in paper II. Also on the basis of the scheme, we show that the circulation qunataization is due to a {\bf circular atomic wave self-interference}. As a whole, in terms of the theory proposed in this work not only are all the explanations regarding the essential properties of He II of a consistent theoretical grounding, but also are consistent with the conceptions that apply to common condensed matter systems and with the common laws of physics. The series of novel concepts evolving from this work can give significant impact also to the understanding of other superfluids.
\end{abstract} 



\section{  Introduction} \label{Secz1}
Liquid $^4$He exhibits properties which are altogether anomalous at temperatures below 2.17 K, i.e. in the  superfluid phase, He II. The two most striking ones are its restraint from solidification down to zero K under ambient pressure\cite{Keesom:1942} and its superfluidity.\cite{WH:Keesom:1930,Allen:Misener:1938,Kapitza,Mehl:etal:1968,JD:Reppy,Henkel:etal:Reppy:1968,allen:jones:1938} Some of the anomalies, such as the large zero point energy\cite{simon} and the volume expansion with falling $T$,\cite{Kamerlingh:1911} can be traced immediately to the extreme chemical character of the $^4$He atom, which being light, small and inert. Many others, on the other hand, cannot be understood in this simple manner; the far-most of these are the superfluidity, the circulation quantization, \cite{vinen,Rayfield:1964,Whitmore-Zimmermann:1968} the $\lambda$ specific heat about the phase transition,\cite{Keesom:etal:1932} the anomalous excitation spectrum,\cite{pol:1958}  and the anomalous thermal conductance\cite{Allen:etal:1937} and melting curve. \cite{Simon:etal:1950} 
These properties can be seen to be largely dictated by the collective and correlation behaviour of the atoms in the condensed form and, their interpretations inevitably require firstly an adequate microscopic depiction of this condensed system. Since the discovery of the superfluidity of He II  during 1911, \cite{Kamerlingh:1911} - 1938 \cite{Allen:Misener:1938,Kapitza,allen:jones:1938} 
extensive investigations have been made on this system both experimentally and theoretically;  although, experimentally direct probe of the excitations using neutron scattering technique, which informs a great deal about the microscopic dynamics,  did not begin until the end of 1950's.\cite{pol:1958} Many comprehensive reviews on this subject haven been given. \cite{london:1954,Gorter:1955,Atkins:1959,Khalatnikov,Egelstaff:1967,Keller:1969,wilks,woods:cowley:1973,Bennemann:1975,brewer,tilleys:86,Glyde:Svensson:1987,nozieres:pines,Donnelly,McClintock:etal:1992,griffin:1993}

Theoretically, the current understanding of He II is essentially based on the compilation of a few distinct models proposed by London,\cite{london:1938} Landau,\cite{landau} and Feynman \cite{feynman:1953} during 1938-1950's; a number of other studies   \cite{tisza,Bogoliubov,cohen:feynman,Pines:1966,lhy:1957,glyde:griffin:90} have also historically contributed to the now accepted "single" model. With one thing in common, all these models  assume that He II consists of two coexisting fluids, the normal fluid and the superfluid, and the two fluids interpenetrate on an atomic scale. The otherwise main concerns in each of the models have been the excitation scheme of the superfluid and a (corresponding) superfluidity mechanism. London \cite{london:1938}$^,$\cite{london:1954} suggested that of the tow fluids, 
the superfluid undergoes Bose-Einstein condensation (BEC), i.e. it consists of atoms that condense in the momentum space into a single degenerate state and in real space are motionless and non-interacting.
Accordingly, the BEC implies a non-viscous and zero-entropy superfluid, and hence yields the superfluidity. 
This is despite the fact that, in his early view, London also foresaw the possibility
 that the quantum states of liquid He II would partly have the feature of the vibration
 states (phonons) of Debye in the theory of metals\cite{Debye}.  
As opposed to London,  Landau \cite{landau} described He II as having two elementary excitations, phonons created mainly at a low $T$, and rotons at a higher $T$. Superfluidity will result if the velocity of a flow motion is lower than that of the particles from the elementary excitations.
A landmark success of Landau's model is the excitation spectrum he phenomenologically designated in order to achieve a desirable (lambda) specific heat; the key feature of the spectrum has been remarkably corroborated, firstly by the neutron scattering experiment of Larsson et al. in 1958.\cite{pol:1958} However, Landau did not provide a microscopic scheme as to how the two elementary excitations occur, and in particular, what actually a "roton" is. Furthermore, for the superfluidity Landau predicted a critical velocity which is at least two orders of magnitude too large than the experimental values, and contains no mechanism for channel width dependence.  The two distinct models above were linked together by Bogoliubov \cite{Bogoliubov} who suggested that the majority of superfluid is in BEC state in which, in contrast to London's BEC, the atoms interact weakly,
and that the transition of the atoms from the BEC state to the quasi particle state produces the two types of elementary excitations of Landau. These two or three models are essentially phenomenological. Feynman \cite{feynman:1953} in the 1950's, in a semi-microscopic description, suggsted the excitations (of both types) result from the superfluid, being the ground state, to the normal fluid. Quantitatively based on variational principle for some trial wavefunction, where the microscopic picture for atomic motion and excitation is however essentially lost,  he gave an improved prediction of the excitation spectrum (1954; Feynman and Cohen,1956). 
At a microscopic scale, Feynman briefly, qualitatively described the atoms as being confined in the cages built by neighboring atoms (-- a picture held also in the present work), but primarily showed the "rotons" of Landau, representing the higher energy excitations, to be possible from a symmetry point of view. 
Taking Landau's critical velocity as a valid start, but to rescue it from the clearly large quantitative discrepancy, Feynman (1955) assumed there present even lower excitations, vortices; the creation of maximum vortices yields critical velocities giving the right magnitudes.
 Until 1990's, the widely adopted (excitation) model of He II is the up-dated compilation of the models by the four individuals, referred to below as the LLBF theory, is as follows: \cite{LLBF} the superfluid corresponds to a BE condensate, which becomes hundred percent at zero \oK; \ the two elementary excitations result from the shuttling of atoms between the superfluid - the BE condensate -- and the normal fluid. Whilst the LLBF model assumes that, at zero \oK, \ the BEC, being identified with  the superfluid, is hundred percent, recent experiments have shown the BEC is only 9 $\sim$ 10 percent. This conflict was circumvented by Glyde and Griffin in the 90's,\cite{glyde:griffin:90} in proposing that the BEC fraction as well as the helium atom mass is modified due to an interaction of the superfluid with the normal fluid atoms.
This apparent improvement does not however remove the basic difficulties pertaining to the foundation of the  LLBF model. Theoretical studies of liquid \He4 \ in the past two decades have otherwise largely involved quantum simulations including variational MC \cite{McMillan,Jastrow,cite-Vitiello:Chester:1990,cite-McFarland:Chester:1994} and path integral MC, \cite{cite-Ceperley:1995,Feynman:1965} which essentially pertain to a more accurate representation of Feynman's (or, the LLBF) excitation model. 
More broadly, these studies are not yet apt to yield the microscopic information of the superfluid atoms, regarding configuration, motion and dynamics, and are even far remote from touching the microscopic scheme of superfluidity, and circulation quantization. 
The Green's function technique for treating many-body problems is not feasible today for treating such complex system as He II, in respect of its above ground propety in parcticular.  \cite{griffin:1993,Woo}   

As a whole, the  LLBF theory encounters various difficulties which give rise to a variety of self-contradictions, and therefore does not represent an ultimately satisfactory theorization for the superfluid He II. Its most significant difficulties may be outlined as follows (further discussions are pointed to each relevant sections).
   1). The  LLBF thermal excitation picture (when faithfully formally represented) inevitably predicts a broad peak in the dynamics structure factor, in direct contradiction to experimental observations (Sec. \ref{Secz2.1}).   
2). The {\it non-viscous, non-interacting} superfluid atoms, which although is sensibly not held in the dynamic model of the LLBF theory but is nevertheless a base for the LLBF explanation of superfluidity -- see item 4) below, directly contradicts with the fact that superfluid atoms are strongly correlated and the actually substantial atomic binding energy (Sec. \ref{Secz2.2}). 
  3). A mixture of two fluids  {\it interpenetrating on an atomic scale} lacks the mechanism for the maintenance of the atomic bonding characters of the respective fluids (Sec. \ref{Secz2.2}).  
4). The critical velocity criteria for superfluidity of Landau \cite{landau} and Feynman \cite{feynman:1953} (1955) have both incorrectly ignored the primary low energy excitations due to phonons (which in fact sensibly presents in the LLBF superfluid dynamic picture), and the comparison of velocities of different masses they made does not conserve energy (Ref. \onlinecite{sfzII}).
\cite{Whitlock:etal:1987}.  
5). The "roton" picture does not predict the  experimentally observed overall features in the excitation at $q_b$ (Sec. \ref{Secz2.2}). 
6). The "Bose-Einstein condensation" picture is itself not capable of explaining superfluidity. \cite{BEC}
These difficulties inherent in the foundation(s) of the  LLBF theory, 
 cannot to us just be removed by a few minor improvements. 
 Therefore, as the motivation for beginning this work, we felt it necessary to develop a theory that starts from an alternative -- an essentially consistent -- theoretical grounding. 

To provide an overall satisfactory interpretation of the complex behavior of He II, we first, in Secs.  \ref{Secz2}-\ref{Secz3}, carry out a systematic derivation of a (microscopic) theory by using, to the broadest possible extent, the key experimental properties of liquid \He4 \  as input information. 
 Three key questions are to be addressed: 1) do the superfluid atoms primarily perform {\it localized oscillations}, or, {\it diffusion}, or do they {\it stand still}  (see Sec. \ref{Secz2.1}) ?   2) whether and how the atoms are interacting (see Secs. \ref{Secz2.2}, \ref{Secz3})?  3) when the two fluids coexist in He II in the higher $T$ end, do the superfluid atoms aggregate in large regions such that they possess a characteristic bonding energy, or, do they intermingle with the normal fluid on an atomic scale (Sec. \ref{Secz2.3})?  The answers to these questions have been achieved in this work after making iterative contrasts between what is predicted by an assumed "theoretical model" and what is observed by experiment, and between the prediction of one property and its implications for others.
The key components of the theory are therefore already characterized in Secs. \ref{Secz2}-\ref{Secz3}. 
In Secs. \ref{Secz4} - \ref{Secz8} we establish the corresponding formal theoretical and quantitative description of He II, including the composition structure of He II (Sec.  \ref{Secz4}), the set-up of equations of motion of the superfluid atoms (Sec. \ref{Secz5}) and solving (\ref{Secz5.2}-\ref{Sec5.4}), and the prediction from the solution of the physical properties, including excitation spectrum (\ref{Secz5.2},\ref{Secz5.4}),  the stability condition of the superfluid (Sec. \ref{Secz6}),  the quantization of circulation (Sec. \ref{Secz7}), and the scattering function near the superfluid bond excitation (Sec. \ref{Secz8}). The superfluidity (its mechanism and the critical velocity) is treated in a separate paper, II. \cite{sfzII}
These above contents  constitute the microscopic part of the theory of superfluid $^4$He. 
 Sections \ref{Secz9}-\ref{Secz10} present a corresponding statistical thermodynamics description of He II. Brief description of the theory has previously been presented in unpublished internal reports  \cite{jane:booklet} except for Ref. \onlinecite{jane:booklet} (2000) which mainly discusses  the experimental probe of the QCE effect as fully presented in paper II.

\section{ The formulation of the microscopic theory based on overall experimental observations} \label{Sec2}  \label{Secz2}

 The static structure factor, $S(q)$, of the (pure) superfluid, as has been probed both by neutrons \cite{Hurst:Henshaw:1955,Henshaw:1960} and by X-ray, \cite{Achter-Meyer:1969,Hallock:1972,Gordon:etal:1958,Keesom:etal:1937} exhibits a heavily damped first peak at about $q_b\approx 1.93$ \AAi and almost vanishing higher rank peaks (FIG. \ref{fig-ex-spec}b). Immediately, this signatures that the superfluid retains only a short range ordering, with a mean atomic separation $a =2\pi/q_b=3.3$ \AA.  \ The detail of this ordering is not our concern; for a large part of  our purpose here it suffices to conveniently, fully represent the atomic configuration with three orthogonal arrays of atomic chains in $X,Y,Z$ directions, we thus have a simple cubic {\it pseudo lattice} with a lattice constant $a$ and a {\it pseudo lattice} reciprocal lattice vector $b=2\pi/a =q_b$; see further Sec. \ref{Secz5.2}.

 Experiments, including specific heat (Secs. \ref{Secz1},\ref{Secz2} ), frequency dispersion  (Sec. \ref{Secz1}), and second sound, \cite{andronik,peshkov,Tam:1987} have commonly shown that He II is an coexistent of the superfluid between about 2.17 to 0.6 ($\sim$1) \oK, \ reduces to predominately a (pure) superfluid below 0.6 \oK,  \ and at zero \oK \ is hundred percent a superfluid. Below in Secs.  \ref{Secz2.1}-\ref{Secz2.3} we infer from experiments the excitation and dynamic schemes in the two respective temperature regions, and the microscopic composition structure of the two-fluid coexistent and the related dynamics.

\subsection{ Thermal excitation via single phonons and atomic dynamics scheme of the superfluid equilibrium } \label{Sec2.1} \label{Sec2.1-PI} \label{Secz2.1} 

\paragraph*{  (i) The dynamic structure factor} \label{Sec2.1.i}  The most revealing indication as to the underlying dynamic scheme in the superfluid equilibrium, which predominates below $\alt 0.6$ \oK, \ is from the dynamic structure factor $S(q,\w)$. The measurements from inelastic neutron scattering \cite{pol:1958,henshaw:w:1961,woods:svensson:1978,Svensson:1987,Stirling:el:1990,Fak-PRB:01,Fak-etal:JdP:00,Fak-PRL:00} have shown that the $S(q,\omega)$ of He II is characteristically different at $T$ below and above $T_{\lambda}$,  at a given momentum transfer $q$, with $q$ raging from 0 up to at least 2.4 \AAi.\cite{sokol}  
 The strongest distinction  presents at an intermediate $q$ ($\sim 1.1 $ \AAi).\cite{woods:svensson:1978,Svensson:1987,Stirling:el:1990}   Here, the $S(q,\w)$  exhibits a superposition of a sharp and a broad peak on the $\omega$ axis, $\omega$ being the neutron energy transfer.\cite{woods:svensson:1978} The broad peak, which has a non-vanishing intensity at $\w=0$, at $T$ above $T_{\lambda}$ identifies with the broad diffuse peak of the normal fluid He I and reduces smoothly in intensity as $T$ is reduced to below $T_{\lambda}$.  The sharp peak appears abruptly as $T$ falls just below $T_{\lambda}$; its intensity increases as $T$ falls and then basically stabilizes from about 1 \oK \ downwards. 
\input epsf  \begin{figure} 
\begin{center} \leavevmode \hbox{%
\epsfxsize= 6.5 cm       
\epsfbox{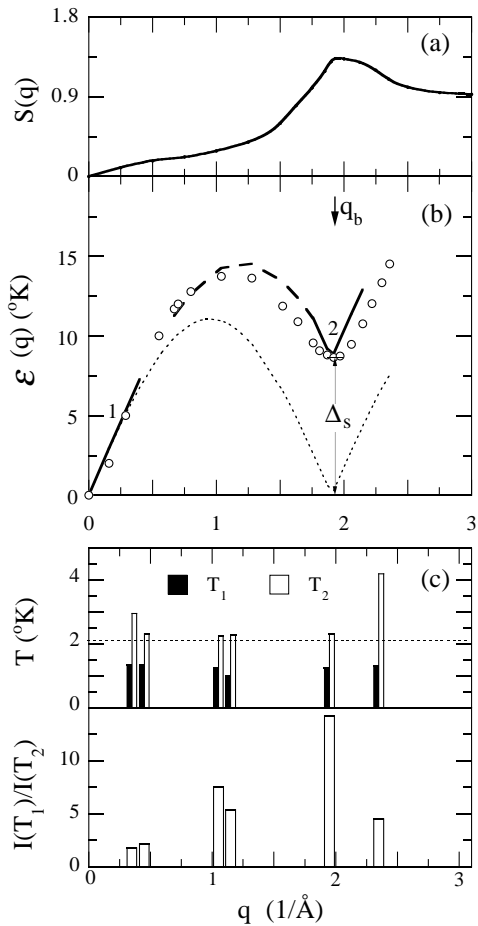}} \end{center}
\caption{   (a). The experimental static structure factor $S(q)$ \ref{Achter-Meyer:1969}
is plotted to assist discussions in the text.  (b). Total thermal excitation energy $\ev(q)$ versus the neutron momentum transfer $q$ (equal to phonon wave number $K$) for the superfluid of He II.  Circles indicate data points from inelastic neutron scattering measurement at 1.12 $\rm{^oK}$\ref{henshaw:w:1961}. The solid and dashed curves show the theoretical dispersion (Sec. \ref{Secz5}); see further FIG. 7.  The dotted curve illustrates a regular (i.e. with no superfluid bond excitation and with no nearest-neighbour repulsion) longitudinal single phonon energy dispersion, given by  Eq. (\ref{eq2-7b}a) for all $K$ values, obtained with $a=3.3$
 \AA \ ($2\pi/a=1.93$ \AA $^{-1}$). 
(c).  Lower graph:  ratio of neutron scattering intensities, $I_1(T_1)$ and $ I_2(T_2)$, at two temperatures $T_1$ ($< T_{\lambda} $) and $T_2$ ($>T_{\lambda}$) as indicated in the upper graph.
\label{fig-ex-spec}\label{figI3}} 
\end{figure}
\noindent 
It is natural to correlate this $T$ dependence of the scattering intensity with  the two-fluid fraction variation just mentioned. Woods and Svensson \cite{woods:svensson:1978} based on quantitative evaluation,  have shown that indeed the sharp and the broad peaks are attributable to the superfluid and the normal fluid respectively. Now, for the superfluid of concern, we can thus concentrate oursevels below only on the sharp peak. The compelling feature with the sharp peak is  that, with the observed finite width being in principle only due to the finite instrumental resolution,\cite{woods:cowley:1973,woods:svensson:1978,Svensson:1987,Svensson:priv} it has the three characteristics of the phonon excitations in a crystalline solid: 
it is qualitatively narrower, it has a high frequency cut-off and, it peaks at a finite $\w$ and has a vanishing intensity at  zero $\w$. Such a sharp peak in $S(q,\w)$ which represents the probability that an energy transfer of $\w$ occurs inevitably implies that, the excitation must have occurred between two discrete energy levels. Experiments using pulse transmission techniques \cite{atkins:stasior} have on the other hand shown that, at long wavelengths, the excitations are via sound waves.  The combined information above hence characterizes the excitation to be via the creation and annihilation of phonons, each having an energy quantum, $\E(q)$.  We refer to these as "single phonons" to separate from the "collective phonons" of Sec. \ref{Secz2.2}. It should be remarked that, phonon excitation, with a well defined atomic dynamic scheme in Debye's theory of solids (1912), \cite{Debye} is terminological also held in the LLBF theory for He II but, as will be illustrated in FIG. \ref{fig-exctyp} below, it clearly is unable to be produced by the dynamics scheme of the LLBF theory.

To infer from the above feature of excitation the underlying atomic dynamic scheme, we shall first be guided by a general principle of quantum mechanics, which states that the state of a quantum particle will be discretized when under spatial confinement. In a bulk superfluid -- a condensed atomic assembly, if and only if assuming the atoms are localized -- in both ends of its excitation, then each atom is wholly confined by its surrounding atoms, as is in a solid. And only then, the atom will then acquire fully discretized states; the transition between these states will result in the sharp peak of the $S(q,\w)$, as is how such a sharp $S(q,\w)$ peak in a crystalline solid (at low $T$) regularly results. (Although, the two present a few important distinctions which we discuss in Sec. \ref{Sec5.2}.)  Based on the above arguments and furthermore on the variety of other uniform indications to be described below, we propose: 
\begin{quotation} \noindent
{\small  In the superfluid equilibrium, \ the thermal excitation is predominantly via the creation and annihilation of (longitudinal) single phonons, whcih predominate at temperatures below 0.6 ($\sim 1$) \oK. \  This excitation can only have occurred \underline{within the (pure) superfluid} and resulted from {\bf localized (harmonic) atomic oscillations} (of an amplitude $\ub(\Rb)$), referred to as the LAO dynamics scheme. Each oscillator roughly corresponds to a single helium atom, separated by the interatomic spacing of He II, $a \simeq 3.3$ \AA, \ and oscillates in a potential well $V$ (of a depth of $\sim -7.2$ \oK/atom as will be detailed in Sec. \ref{Secz2.3}).} \end{quotation}

Based on the LAO dynamics scheme, the solution of equations of motion of the superfluid atoms in Sec. \ref{Secz5} actually reproduces the experimentally observed excitation feature, i.e. the excitation is between discretized energy levels $\e_n$, $n=0,1,\ldots$ (FIG. \ref{fig-exctyp}a), each differing by an energy quantum of a phonon, $\E(K)$. In effect, LAO is the only dynamic scheme able to give rise to a sharp diffraction peak for the superfluid. In contrast, the LLBF picture of phonon excitation  (FIG. \ref{fig-exctyp}b), namely {\it the excitation results from the knocking out of atoms from the Bose-Einstein condensate into the normal fluid}, will incorrectly predict  a broad peak of the dynamic structure factor. Evidently, this is because the energy of an atom in the normal fluid end, $E(p)$ at a given momentum $p$, and consequently the excitation energy $E(p) - E_{BEC}$, are continuous, despite the single valued BEC energy $E_{BEC}$. 

\input epsf  \begin{figure} [here]
\begin{center} \leavevmode \hbox{
\epsfxsize= 9.5 cm        
\epsfbox{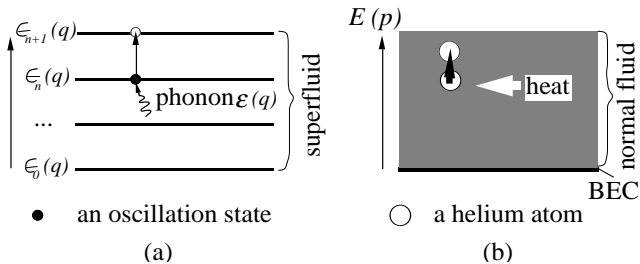}} \end{center}
\caption{ 
Excitation scheme for the superfluid equilibrium as proposed: (a) by the present theory.
The excitation is via the creation of a single phonon, corresponding to a transition between the discrete energy levels $\e_n(q)$ and $\e_{n+1} (q)$ of the oscillation mode $q$ of the localized (relative to $\xi$) superfluid atoms.  (b) by the LLBF theory. The excitation of the helium atomic motion is produced by a transition from the single degenerate state of the "Bose-Einstein condensate" (the superfluid) of energy $E_{BEC}$, to the continuous normal-fluid energy $E(p)$ at a given momentum $p$. 
\label{fig-exctyp} \label{figI1} } 
\end{figure}

The inference of localization and the LAO from the sharp $S(q,\w)$ peak is further supported by the observation of a one-to-one link between a sharp $S(q,\omega)$ (or broad) peak and atomic localization (or diffusion)  which being general and system-independent, as has been identified, for the first instance, by Sk\"old and Larsson \cite{Skold:Larsson:1967} in their comparison of the scattering features from the solid and the classical liquid phase of argon, and similarly from the two different phases of other systems. Where, the broad scattering intensity at lower energies was assigned to be due to diffusion, the so called "quasielastic scattering".  In another example, a sharp $S(q, \omega) $ peak centered at a finite $\omega$ for a fixed (intermediate) $q$ has also been observed in various metallic liquids, such as lead by Egelstaff \cite{Egelstaff:1966} and rubidium by Copley et al. \cite{Copley:Rowe:1974}  Copley and Lovesey \cite{copley:lovesey} correlated that feature to the stronger metallic bonding of Rb compared to that of a classical liquid. We can readily infer that the strong bonding functions to confine -- to localize -- the atomic motion. 


The LAO scheme also immediately explains as to why mainly the intermediate $q$ scattering rather than the low $q$ is specific with the superfluid; at a $q$ $<0.4$ \AA$^{-1}$, for $T$ below and above $T_{\lambda}$, the position, intensity and width of the $S(q,\omega)$ peak of liquid helium exhibit no distinct changes except that the rate of increase of the width is slowed down. \cite{Stirling:el:1990}
Since $K$ ($\equiv q$)  $\sim 1.1$ \AAi corresponds to a wavelength $\la$ ($=2\pi/ q$) $\sim$ 10 \AA \ crossing over only a few atomic spacings (with $a \simeq 3.3$ \AA).
Hence a coherent oscillation, if to occur at such a short wavelength, would require all of the atoms move cooperatively and the atoms must therefore be localized. By contrast, at long wavelength crossing over many atoms, the oscillation occurring by means of density fluctuation is insensitive to diffusion of some individual atoms. Indeed, in addition to the observation for liquid helium above, neutron scattering experiments have shown that the $S(q,\w)$ functions from normal liquids all present a certain sharp peak at $q\alt$ 0.4 \AA$^{-1}$. \cite{copley:lovesey}  
 

\paragraph*{  (ii). The excitation spectrum }\label{Secz2.1.ii} \label{Sec2.1.ii}  
The LAO scheme is further pointed to by the excitation spectrum (FIG. \ref{fig-ex-spec}b), $\omega(q)$ due to the sharp peak of $S(q,\w)$  \cite{henshaw:w:1961}  in two respects.  First, except for the peculiar non-zero value, $\Delta_s \simeq $8.6 \oK \ at $q_b \simeq$1.93 \AA$^{-1}$, \ $\omega(q)$ has a sinusoidal dispersion  which is the solution of the equation of motion of (relatively) localized helium atoms, Eq. \ref{eq2-7b} in Sec. \ref{Sec5.2}, for the longitudinal mode. Second, the solution informs  the linear sector to be due to propagating elastic sound waves, here phonon waves, of a velocity $ 238.9$ m/sec which is in near exact agreement with the first sound velocity $c_1 =  239$ m/sec from pulse transmission experiments.\cite{atkins:stasior}  Moreover, the experiment by Woods showed that the slope here is independent on $T$, \cite{Woods:1965} and disproves the postulation \cite{Hohenberg:1964} that the slope might be instead due to the two-fluid density variation with $T$ ($c_1 (\rho_s/\rho)^{1/2}$).  Furthermore, the dispersion "period", $q_b = 2\pi/a \simeq1.93 $\AA$^{-1}$, giving $a \simeq 3.3$ informs the oscillator size $a_{\rho} \sim 3.6$ \AA,\cite{a} \ which is just a helium atom. This can be similarly seen from the upper limit, $q \simeq 0.6$ \AA$^{-1}$, of the propagating phonon wave which development would require at least be a few oscillators within a wavelength. In contrast, the part of the "in-phase" oscillation of a large ($N_M$) number of superfluid atoms (to be discussed in the next section), which alone is similar to the weekly interacting BEC wave or the zero sound wave \cite{Pines:1966} in the LLBF terminology, has a maximum dispersion period $K_M \propto 2\pi/N_Ma << q_b$ and an excitation energy negligibly low compared to the one of the single phonon here.

\begin{figure} [here]
\begin{center} \leavevmode \hbox{%
\epsfxsize= 9 cm     \epsfbox{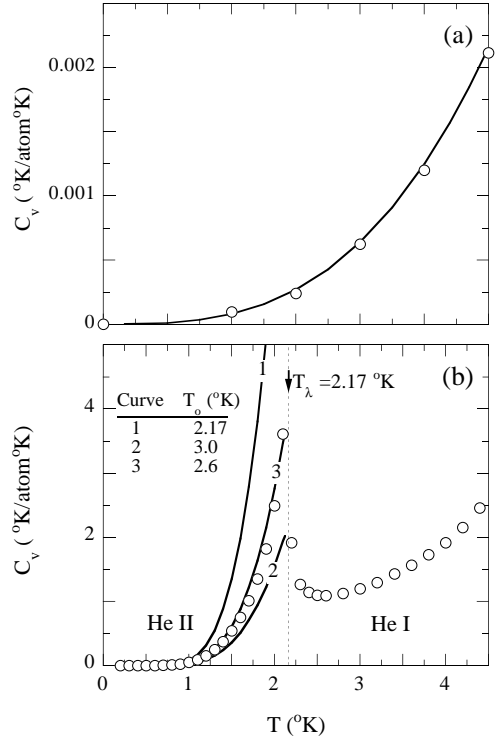}} \end{center} 
 \caption{ Specific heat of He II, $C_V(T)$.  Circles represent experimental values (Refs. \ref{wiebes:el,Kramers:1952,hill}); solid lines are statistic thermodynamic results of Sec. \ref{Secz9} as based on the present microscopic theory of Secs. \ref{Secz2}-\ref{Secz5}. 
 (a) for $T$ below 0.6 \oK, the theoretical $C_V(T)$ $\propto T^3$, Eq. (\ref{eq3-31})$'$, is given by the derivative of the internal energy  obtained from a canonical ensemble average over the states, with a Debye type of density of states, of the independent, indistinguishable single phonons, which are produced due to in real space the harmonic oscillations in the normal coordinates of the ($N$) independent, localized superfluid atoms relative to $\xi$. (b) from 0.6 to just below 2.17 \oK, \ the theoretical $\lambda$-$C_V(T)$ (Eq. (\ref{eq3-31})$'$) results principally from the conversion of the superfluid to the normal fluid, via the excitations of superfluid bonds. 
The lower and upper limits of $T_o$, $2.17 $ and $3.0 \rm{^oK}$, yield the uncertainty boundaries of the theoretical $C_V(T)$ shown by the solid lines 1 and 2; $T_o=2.6$ \oK \ gives the optimum $C_V(T)$, solid line 3.   \label{fig-Cv} }  
 \end{figure}

\paragraph*{  (iii). The $T^3$ specific heat} \label{Secz2.1.iii}  The LAO scheme is indicated by the Debye $T^3$ behavior of the specific heat $C_V$ of He II between 0 $\sim$ 0.6 \oK \ (FIG. \ref{fig-Cv}a). \cite{wiebes:el} 
Insofar as a non-conducting condensed atomic system is in question, a Debye $T^3$- $C_V$ is broadly known to be necessarily given rise to by an atomic assembly where the atoms are localized and perform harmonic oscillations; the superfluid is not an exception as Sec. \ref{Secz9.3} explicitly treats based on statistical thermodynamics. By contrast, a classical fluid consisting of diffusive atoms generally shows a Petit-Dulong heat capacity. It follows that, the LLBF excitation scheme will inevitably yield a specific heat strongly bearing the feature of the normal fluid, which cannot be of a $T^3$ behavior. Furthermore, the theoretical Debye $C_V (T)$ obtained in Sec. \ref{Secz9.3}, when fitted to the experimental $C_V$ data, produces a longitudinal phonon velocity of $ 241$ m/sec. This value agrees (within experimental error) with the $c_1 $ value (239 m/sec) from pulse transmission \cite{atkins:stasior} as well as phonon velocity (238.9 m/sec) from the excitation spectrum discussed in (ii).  
\input epsf  

\paragraph*{ (iv). The inferred diffusion rate }  \label{Sec2.1.iv}  The LAO scheme is directly supported by the solid-like low diffusion rate in He II, being $10^{-8}$ cm$^2$/sec as Eq. \ref{eq-dif} in Sec. \ref{Secz5.4} shows. The associated diffusion barrier implies a negative potential well in which the atom executes oscillation.   

\paragraph*{ (v).  The capability of predicting superfluidity} \label{Sec2.1.v} The preceding observations inform that the superfluid atoms are not {\it standing-still} (in either the absolute or relative coordinates), and rather, the superfluid is fully thermally excited, with the excitation states, as Sec. \ref{Secz9} shows, exhibiting a Planck distribution. The LAO scheme and the implied property above are furthermore supported by the fact that, as paper II shows,  on the basis of the LAO scheme the QCE superfluidity mechanism can consistently, quantitatively predict superfluidity and critical velocity both in satisfactory agreements with experimental data.

The one-to-one connection between an excitation spectrum as FIG. \ref{fig-ex-spec}b for He II and a fully excited He II is an inevitable truth, if one compares it with the analogous excitation phenomena of the vast body of crystalline solids, in terms of the phonon feature in general and the $Kc_1$ behaviour in the linear sector in particular. Had Landau's criterion of critical velocity  for He II,  namely that if $v_s < c_1$ there is no phonon excitation, reflected the truth, then one will be led to the inevitably untrue conclusion,  that any piece of crystalline solid moving, say against the surface of the ground, with a speed less than its sound velocity would be in superfluidity motion.

\subsection{The superfluid bond excitation and atomic dynamics scheme; the gradual  phase transition of the superfluid  } \label{Secz2.2} \label{Sec2.3} \label{Sec2.3-PI}  
\begin{figure} [here]
\begin{center} \leavevmode \hbox{%
\epsfxsize= 6.5 cm     \epsfbox{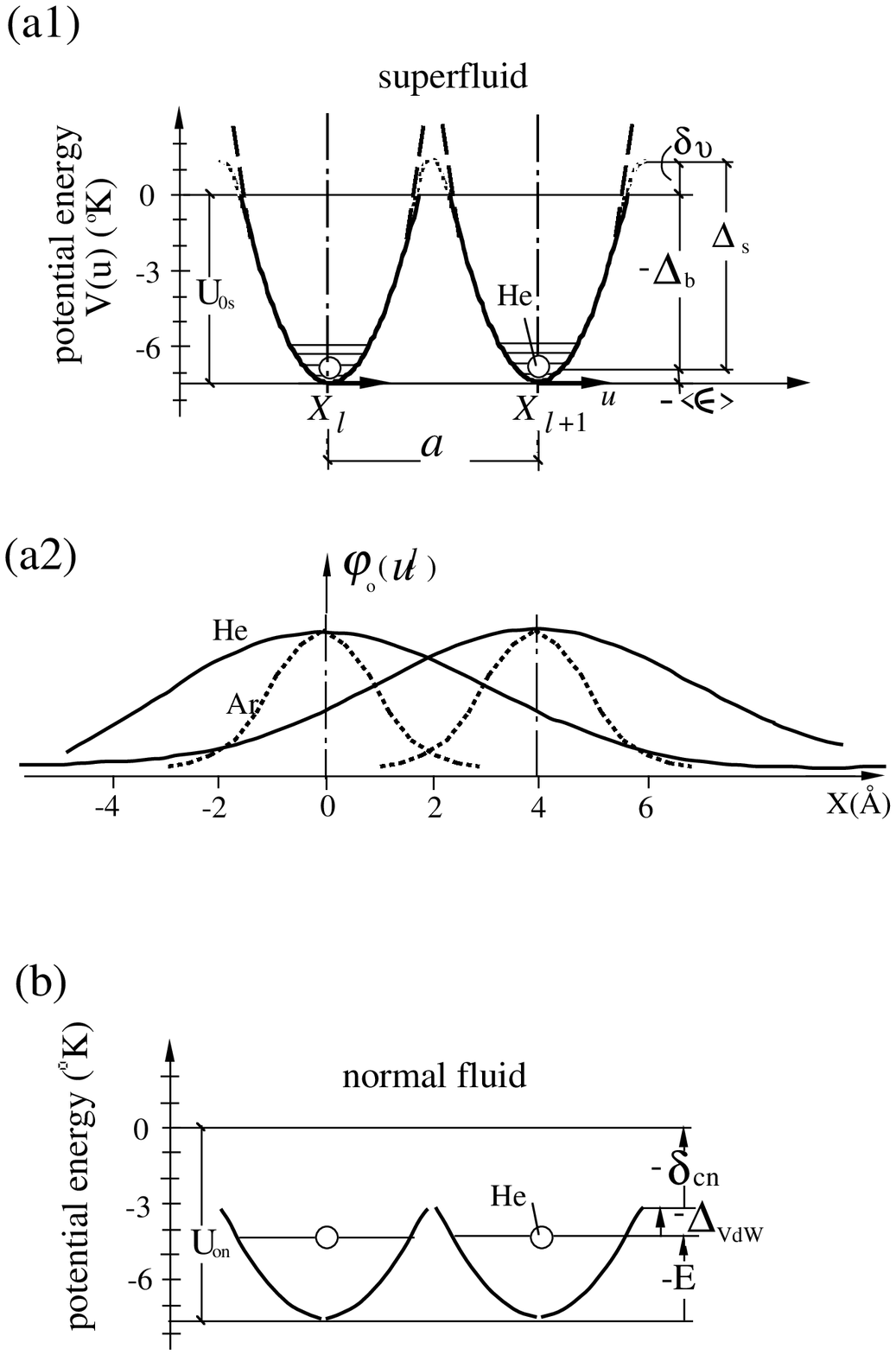}} \end{center} 
 \caption{   (a1). The potential energy $V(u)$ for a superfluid atom at $X_l$ and for one separated at a distance $a$ at $X_{l+1}$, solid parabolic curves, theoretically constructed as in Eq. (\ref{eq:pot:1})$"$ illustrated for the mode $Ka=0.5 \pi$; see FIG. \ref{fig-Vu} for $V(u)$ at different $K$ values. 
In the potential wells, the superfluid atoms perform independent harmonic oscillations in the relative coordinates $\xi$.
The potential well depth $V_0$ is given by the experimental ground state energy $U_{0s}=-7.2$ \oK/atom \ref{simon:el:1950}. The activation of an atom from the bottom of the well to just above zero, at a energy cost  $\Ds=\Db+ \delta_v$, represents the excitation of a superfluid bond $\Db$, with a site creation energy $\delta_v$. $<\e >$ represents the average kinetic energy per superfluid atom, and by Sec. \ref{Secz9} $<\e > <<U_{0s}$, hence $\Delta_b \simeq U_{0s}$. 
The dashed lines indicate the repulsive  part of the potential, $V(u)>0$, and is associated with the phonon excitation within $0.6 \le K \le 1.4$ (1/\AA) \ where $\E(K) > \Delta_b$; cf. FIG. \ref{fig-ex-spec}.  (a2) The ground state wavefunctions (solid curves) resulting from the solution of equation of motion of superfluid atoms. The dotted curves show wavefunctions for Ar atomic mass. (b). Illustration of the normal fluid potential energy for two adjacent atoms.  $E=U_n-U_{on} \simeq 3.1 $ \oK \ for $T$ just below $T_{\lambda}$ is the kinetic energy, and $\Delta_{VdW}$ the average Van der Waals bonding energy.  The diffusion barrier height between the adjacent atoms, being about 1.6 \oK \ relative to $E$, is estimated such that it produces a diffusion constant in the order of $10^{-4}$ cm$^2$/sec at 1 \oK, a typical value for liquids. $V=-\delta_{cn}$ is the energy region in which the atoms are free atoms throughout the bulk.  $U_{on}=U_{os}$ is assumed.  \label{fig-harmpot2}  }  \end{figure}

The atomic localization in the LAO dynamics scheme derived in Sec \ref{Secz2} immediately suggests that, in the superfluid the atomic bonding has undergone a qualitative enhancement compared to the van der Waals (vdW) "bonding" energy of the normal fluid He I. We represent the former by {\bf superfluid bond}, $\Delta_b$, which is defined to be the negative of the binding energy per superfluid atom that is given rise to by the attraction from many superfluid atoms within their correlation regime, $\Up$. Clearly, $\Delta_b$ is directly associated with the stability of the superfluid phase. We propose:

\begin{quotation}\noindent {\small A {\it quasi equilibrium} process is predominated by excitations of superfluid bond ($\Db$) in two closely related $\Db$ dynamics schemes, where $\Delta_b$ represents an attractive energy barrier which prevents the reference atom 1) from diffusing into its neighbouring site within the superfluid when locally perturbed, due to thermal fluctuation or an external incidence such as a neutron, and,  2) from transforming into the normal fluid state when globally perturbed, e.g. by bulk heating-up between 0.6 $\sim$2.17 \oK. } \end{quotation}
\noindent
By the definition above, $\Delta_b$ can be expressed by 
\begin{eqnarray} \label{eq-Db} \Db 
 = - \Us0 -<\e_n> + \us. \end{eqnarray}
 Where, $\Us0$ is the ground-state cohesive energy (per atom) of the superfluid, representing the depth of the potential well in which a He atom is confined (FIG. \ref{fig-harmpot2}a1);  $U_{s0}=- 7.2$ \oK/atom as given by thermal measurements at SVP. \cite{simon:el:1950,london:1954} $<\e_n>$ is the per-atom kinetic energy of superfluid atoms and is $<<\Us0$ as will be verified in Sec. \ref{Secz9}. $\us (<0)$ represents the "long range" attraction energy by atoms in the entire fluid bulk, so an atom in ($\us$, 0) can diffuse virtually freely. The superfluid atom is, as Sec. \ref{Secz5.1} will show, localized with respect to a kinetic energy even up to $- \Us0 $, thus one can expect $\us << \Db$; hence we have 
$$     \Db \approx - U_{s0}= 7.2 \quad  {\rm (^oK)}. \eqno(\ref{eq-Db})^{\prime} $$
\input epsf  
Cases (i)-(iii) below show further experimental indications of the superfluid bond.

\paragraph*{ (i). The $\la$ specific heat} The superfluid bond excitation upon a global perturbation is directly indicated by the experimentally observed $\lambda$- $C_V$ of He II between about 0.6 to 2.17 \oK \ (circles, FIG. \ref{fig-Cv}b). \cite{Kramers:1952,hill}  
For if the superfluid to normal fluid conversion, which occurs in large quantity here, energetically involves the conversion of $\Db$ to the qualitatively smaller vdW bond of the normal fluid (FIG. \ref{fig-harmpot2}a,b), then inevitably a large excess of heat supply is required. The activation energy, denoted by $\Delta_{sn}$, is thus consumed mainly for $\Db$ and, additionally for a site creation energy $\dn$ and the kinetic energy $E$ of a normal fluid He atom. Hence, 

\refstepcounter{equation} $$ \label{eq1-1} 
\Dsn = \Db + \dn + E.   \eqno(\ref{eq1-1}a)$$
Using (\ref{eq1-1}a), the thermodynamic evaluation in Sec. \ref{Secz9} will justify the excitation scheme above to actually reproduce the experimentally observed $\lambda$-$C_V$:  $\propto \exp(-{\Dsn \ov k_BT})$.   

Furthermore, the $\Db$ excitation scheme is supported by the observation that, a qualitative bonding change when transforming from one phase to the other is the common underlying excitation scheme for the second order phase transitions of a variety of systems (e.g. alloy) with a $\lambda$ specific heat.

\paragraph*{(ii). The dynamic structure factor about $q_b$ } The superfluid bond excitation upon a local perturbation is directly pointed to, in the first instance, by the peculiar features in the $S(q_b,\omega)$, at $q_b \simeq 1.93$ \AA$^{-1}$, of the superfluid (FIG. \ref{fig-ex-spec}b).\cite{pol:1958,henshaw:w:1961,Stirling:el:1990,Talbot:el:1988,svensson:el:1996} Two of these are: 1) it peaks at a finite $\w$, $\w(q_b) = \Delta_s \simeq 8.6$ \oK \ which is basically independent of $T$.
  2) its peak width is qualitatively narrow and on the $T$ axis exhibits a sharp edge at $T_{\lambda}$.\cite{henshaw:w:1961} By contrast, the peak of $S(q,\omega)$ of He I is broad, and its position diminishes towards zero as $T$ rises, e.g. to $\sim$4.5 \oK/atom at $T=2.9$ \oK. \cite{Stirling:el:1990} On the other hand, a simple cubic crystalline solid, for example, would show a zero frequency at $q_b$. The underlying $\Db$ dynamics for the above is explained below.
 At $q_b$, which reciprocal being $\simeq 3.3 = a$ \AA \ as noted earlier, the solution for the equation of motion  (Eq. \ref{eq7}) relative to $\xib$, the dotted line in FIG. \ref{fig-ex-spec}b,  gives no phonon excitation. And, in real space, as viewed by a neutron committing a momentum exchange of $q_b=K_b$, all atoms are located at the bottoms of their potential wells;  cf. Sec. \ref{Sec5.2}(ii). Hence, when a He atom here is knocked by a neutron transferring an energy $\Ds$ just a fraction larger than the well depth $|\Us0| \sim \Delta_b$ (FIG. \ref{fig-harmpot2}a),  it is, if moved at all, knocked to just a fraction above the well. Then, the $\Delta_s$ is mainly consumed to excite the $\Delta_b$ and in addition, for a small site creation energy $\ds$. As similar to Eq. (\ref{eq1-1}a), we therefore have: 
$$ \Ds = \Db + \ds.   \eqno(\ref{eq1-1}b) $$
As the superfluid atoms are localized, $-\Delta_b$ and $\ds$ are therefore relatively sharply defined except for a finite smearing, primarily in $\ds$, due to the short-range-ordering resultant fluctuations in the instantaneous atomic configuration. This therefore necessarily explains the qualitatively narrower peak in $S(q,\w)$. On the other hand, a finite smearing implied above is indeed supported by neutron scattering experiments which have shown the $S(q,\w)$ peak -- though narrow -- has a finite width (e.g.  systematically presented in FIG. 8 of Ref. \onlinecite{henshaw:w:1961}), being a couple of times the instrumental resolution as used e.g. in Ref. \onlinecite{Fak-PRL:00}, and being also at least several times wider than Landau's theoretical prediction based on roton-roton collisions. In contrast, in a normal fluid the essentially free atomic diffusion results in a significantly broad peak in the $S(q,\w)$.

The $\Db$ dynamic scheme is further supported by the third feature in the $S(q_b,\w)$, 
namely that its intensity is the highest compared to at other $q$ values. From the experimental data of Refs. \onlinecite{Stirling:el:1990},\onlinecite{woods:svensson:1978}, the relative intensity, 
i.e. the ratio of the intensities, $I_1$ and $I_2$, below and above \Tlmb\ respectively is estimated to be 14 at $q_b$ (FIG. \ref{fig-ex-spec}c), which is about two times the second highest intensity, $I_1/I_2 \simeq 7$, at $q=1.1$ \AA$^{-1}$. 
An amplification of the scattering amplitude at $q_b$ is to be expected, if an incident  beam of  monochromatic neutrons are indeed scattered by single atoms from an array of plans at a mean separation $a$,  at which the condition $q_b=2\pi/a$ implies an effective  Bragg scattering (see further Sec. \ref{Sec8}). 

That the same dynamics underlies both the $\lambda$ specific heat and the excitation at $q_b$, as implied in (i) and (ii), can be justified by the fact that the latter indeed constitutes the major contribution to the former, as is shown both by our thermodynamic result, Sec. \ref{Secz9}, and by the accurate numerical calculations of Yarnell \cite{ypb:1959} and others. \cite{Donnelly:etal:1981} It is noteworthy that, a frequency dispersion with such a minimum at $q_b$ was initially devised by Landau in his well known 1941 paper, in order to yield a $\lambda$ specific heat. We here see that, the correspondence between  the $C_V$ and the $\w(q)$ does not necessarily imply "which" dynamics.

 \paragraph*{(iii). The ion mobility } The $\Db$ dynamics scheme upon local perturbation is indicated, in the second instance, by the experimentally observed behaviour of the He ion mobility $\mu$ in He II,\cite{Careri:el:1959,Reif:el:1960,McClintock:etal:1985} which obeys 
\begin{eqnarray} \label{eq:ion} \mu =  v_i / |{\bf E}|_a \propto  \exp(\Delta /k_B  T), \end{eqnarray}
${\bf E}_a$ being the applied field.
Since by definition mobility is inversely proportional to the ion scattering rate and hence to the excitation probability $P \propto \exp({- \Delta_s /k_BT})$ (Sec. \ref{Secz5.3}):  namely it is $ \propto 1/P \propto \exp({ \Delta_s /k_BT})$, which is just as represented by Eq. (\ref{eq:ion}). Hence the $\Delta$ of (\ref{eq:ion}) identifies with the excitation energy, which is determined in the experiments to be 8.1 \oK \ (for negative ions) or 8.8 \oK \ (for positive ions).  $v_i$ is the ion drift velocity, having a magnitude typically 10 $\sim $ 100 m/sec.
Clearly, the ion has a much higher kinetic energy  than the thermal drift motion of a helium atom which  velocity is in the order of $10^{-10}$ m/sec (see Sec. \ref{Sec5.4}).
%
Hence in the ion drifting process, the ion is essentially not subject to a many-quantum-atom correlation;  and also the ion principally will transfer its energy and momentum to the superfluid, similar to the thermal neutron scattering. Hence, as discussed in (ii), the scattering is due to the $\Delta_b$ excitation. Indeed, the basically equality of $\Delta$ ($\simeq 8.1 \sim 8.6$ \oK) from the ion scattering and the $\Delta_s$ of neutron scattering (Sec \ref{Sec2.3}(ii)) further supports the $\Db$ dynamics scheme.

The excitations in the two environments discussed in (i) and (ii)-(iii) can be generalized to be
$$  \Dsa \simeq \Db + \da + E_{\alpha}, \eqno(\ref{eq1-1}) $$
$\alpha$ omitted for the local perturbation, $\alpha=s$, and $n$ for the local and the massive bond break-up, $E_{\alpha} = <\epsilon> \simeq 0$ is the thermal kinetic energy of a He atom in the superfluid, and $E_{\alpha} = E$ in the normal fluid of He II.  
Using in (\ref{eq1-1}b) the known values of $\Delta_s$ and $\Delta_b$ and with $E_{s} \simeq 0$ gives $\ds = 8.6-7.2 =1.4$ \oK. 

\subsection{ The composition structure of He II, the momenton wave and the collective phonons} \label{Sec2.2} \label{Sec2.2-PI}  \label{Secz2.3} 

\begin{figure} [here] \begin{center} \leavevmode \hbox{%
\epsfxsize= 9. cm     \epsfbox{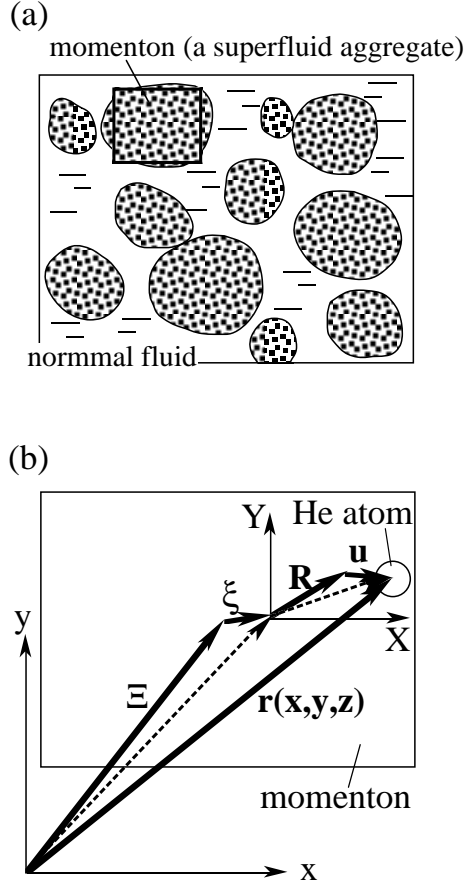}} \end{center} 
\caption{ (a). A snapshot of the schematic composition structure of He II at a $T$ at which both fluids have a significant fraction. Each superfluid aggregate, a dotted region, represents a momenton; the momentons are randomly distributed through a normal fluid background. (b). In a momenton region, out of  the thick framed region in (a), 
the absolute coordinates $\rb$ of a superfluid atom is represented by the vector sum of four relative coordinates: $\ub$ describes the atomic oscillation relative to $\Rb$,  $\Rb$ the translation of the equilibrium position of the atom viewed from $\xi$, the coordinates of the center-of-mass of the momenton, $\xi$ the oscillation of the momenton relative to $\Xib$, and $\Xib$ the translation of the equilibrium position of the momenton.   \label{fig-mom}}  \end{figure}

The realization of the relative localization of the superfluid atoms  (Sec. \ref{Secz2.1}), and the formation and preservation of their large bonding (to be justified in Sec. \ref{Secz2.2}) and its many-atom correlation origin (Sec. \ref{Secz3}), all point to a composition structure of the two-fluid coexistent as described with a {\bf momenton} model (FIG. \ref{fig-mom}): 
\begin{quotation} \noindent
{\small A {\bf momenton} is firstly an aggregation of a large number of superfluid atoms, of an irregular shape and size of macroscopic scale. The motions of the atoms in it are primarily the $\ub(\Rb)$- oscillations  as described in Sec. \ref{Sec2.1}, which are about sites $\Rb$'s fixed in the center-of-mass coordinates of the momenton, $\xib$, here. The two-fluid coexistent of He II consists of tremendously many momentons adjacent to the normal fluid regions, or within large (pure) superfluid regions.
Dynamically as a whole, in a He II in thermal equilibrium at rest, each momenton oscillates slowly with an amplitude $\xib(\Xib)$, about its equilibrium position, $\Xib$; this motion  is equivalent to the {\it simultaneous} and {\it in-phase} oscillations of all of the atoms in it, which (being of a small proportion) are superposed to the prominent coherent, not-in-phase $\ub$ oscillations of the single atoms  of Sec. \ref{Sec2.1}. 
Under a pressure difference, $\Xib$ describes the translation of a momenton, 
and hence the steady flow motion in the absence of convection. Finally, the excitation of  a momenton occurs via the creation/annihilation of {\bf collective phonons}, giving rise to the {\bf momenton waves}.} \end{quotation}

\input epsf  

 In contrast, {\it a mixture of two fluids which are intermingled on an atomic scale,} as is assumed in all of the existing models (or, the LLBF theory), lacks the mechanism for the preservation of the interaction of the respective fluids. The formation of momentons also is energetically favourable in two ways: 1) given the large superfluid atomic binding originates from a many-quantum-atom correlation (Sec. \ref{Secz3}), the gathering together of the superfluid atoms will allow an optimum atomic attraction, thus yielding the lowest energy of the system. 2) it requires less energy to turn a normal fluid atom into a superfluid atom, if the former is at the edge of the existing superfluid region rather than inside the normal fluid. 
Furthermore, based on the momenton model of composite structure, the two-fluid fractions can be quantitatively predicted with only the experimental density of liquid helium as an input data (Sec. \ref{Secz4}), in satisfactory agreement with the statistical thermodynamics treatment in Sec. \ref{Secz9}.  

Direct experimental indications of the momenton wave include:  (i) A slow propagation of the elastic wave -- the so-called second sound wave -- has been observed, e.g. by Peshkov et al. \cite{peshkov} and Atkins et al., \cite{Atkins:etal:1950}  in thermal pulse transmission measurements of He II.  It occurs only at low frequency (less than $ 1000 $  cycle/sec); it has a slow propagation velocity, $c_2 \simeq $ 20 m/sec within $2.17 \sim 1$ \oK, \ and from about 1 \oK \ downwards, $c_2$ increases rapidly and the pulse becomes substantially broadened. Despite the broadening, a finite width of the pulse is observed even with $T$ being as low as 0.7 \oK, \ which tends to suggest a finite size of the momenton in the superfluid. (ii). The thermal conductivity in He II  \cite{Allen:etal:1937} is observed to be abnormally large and not defined by a temperature gradient.  The momenton wave essentially describes the phenomena of (i)-(ii).  For instance, the simultaneous and in-phase oscillation of many atoms (of a momenton) in effect will produce the instant propagation of a heat pulse; in other words, the pulse is at once carried forward by a distance having the dimension of a momenton. 

\section{ The origin of the superfluid bond}  \label{Secz3} 

The experimental evidence of Secs. \ref{Secz2.1}-\ref{Secz2.2} together points to the cooccurrence of the large atomic bonding and the (relative) localization of the atoms in the superfluid. Such a cooccurrence commonly takes place in such processes as liquid solidification, for which the physics is generally understood. On the other hand, the physics for the superfluid atomic localization, 
which occurs uniquely as a result of the transformation from the classical state He I to the quantum fluid He II,  appears specific.
From the basic fact that the superfluid atoms are quantum mechanical, we can infer their enhanced bonding mechanism as follows. A quantum (fluid) atom is marked by a broadly spread wavefunction $\psi(\rb)$, through a volume at least a few times greater than $a$ (see Sec. \ref{Secz5}).  In this volume, at any location we will find a fractional density of atom $A$, $\psi_A^2(r_{A_i})$ and that of atom $B$, $\psi_B^2(r_{B_j})$; for simplicity, we call them partial-atom $A_i$ and partial-atom $B_j$.  Each partial atom consists of a fractional nucleus (of a charge $\psi^2(r) Z^{2+} $) and a fractional electron cloud (of a charge $\psi^2(r) Z^{2-} $) in just the way as the complete He atom, except for each being weighted by the fractional density $\psi^2(r)$ ($\int_{V} \psi^2(r) d \rb = 1$.) The above yield three characteristics in the atomic interaction.  1). the exchange-correlation is amongst all of the partial-electrons of all of the partial-atoms, as well as among all of the partial nuclei, that tend to approach the same locations.  In this complicated correlation process, the surrounding atoms extending to many neighbors are strongly involved with one another. 2). given an $AB$ atom pair (meaning $A$ sitting to the left of $B$), some of the partial-atom pairs can assume the $B_j A_i$ configurations.  These $B_j A_i$ pairs will repel with each other when $A$ and $B$ are tending to part.  Both of the above aspects will lead to the observation that the atoms are prohibited from separating from one another and are thus driven into a (relative) localization.   3). the interaction strength between a pair of partial-atoms is only a fraction of that of the total $AB$ interaction.  This will particularly affect, actually soften, the short-range repulsion.

In addition, the localization allows a cooperative motion of many helium atoms, the LAO scheme.
 The atomic polarizations are therefore induced systematically, which essentially corresponds to the London or dispersion energy, \cite{London:1930,Seitz:1940} of the form $e^4h\nu_0/2\alpha^2r^6$, rather than randomly as with the normal fluid van der Waals interaction.

\section{  Quantitative description of the composition structure of H\lowercase{e} II, the two fluid fractions  }\label{Secz4}  \label{Sec4} \label{PSec2} 

Consider a He II bulk at a given $T$ of the composite structure as depicted in Sec. \ref{Secz2.3}; let the $\imath$th superfluid ($s$) or normal fluid ($n$) region in it be denoted by $V_{si}$ or $V_{ni}$. Then one expects that the $s$ and $n$ component each has its own specific density, denoted by $\rho_s$ and $\rho_n$.
Incorporating further the factor that the atoms of the normal fluid are classical and of the superfluid are quantum mechanical, the local density of He II, i.e. the number of atoms per unit volume at $\rb$ is thus
\begin{eqnarray} \label{eq2a}          
 \rho(\rb, T) =          
 \left\{ \begin{array}{cc} 
  \rho_s(\rb, T) =  \sum_{j}  | \psi_s(\rb_{j})  |^2,    
& \mbox{$  \rb_j, \rb \in V_{si}$}, \cr  
\rho_n(\rb, T) =  \rho_n \delta ( \rb_j),  
& \mbox{$  \rb_j, \rb \in V_{ni}  $},  \end{array}  \right.       
\end{eqnarray}
or,
$$ \rho(\rb,T) = \rho_s(\rb, T) \delta (\rb \in V_{si}) + \rho_n(\rb, T) \delta (\rb \in  V_{ni}).  \eqno{(\ref{eq2a})'} $$
where $\psi(\rb_j)$ and $\delta(\rb_j)$ are the wavefunctions of a quantum and a normal helium atom respectively. The $\sum_j$ runs over all the atoms within the wavefunction overlap region including $\rb_j = \rb$. 

We can define the volume fractions of the $s$ and $n$ fluids respectively by
\begin{eqnarray}  \label{eq2}
& & f_s(T) = { \sum_i V_{si}  \over   \sum_{i} V_{si} +  \sum_i V_{ni} } = { V_s \over V }, \qquad {\rm and}\\ \label{eq3} 
& & f_n(T) = 1 - f_s(T)  = { V_n \over V },  \end{eqnarray}
where the sums cover all of the regions of the respective fluids in the container, and $V$, $V_s$, and $V_n$ are the total volumes of the two-fluid mixture, the $s$ and the $n$ fluid, respectively.  The average density of the two-fluid mixture is therefore 
\begin{eqnarray}  \label{eq4} 
 & \rho(T) & = (1/V)\int_V  \rho(\rb, T) d \rb  \nonumber \\
  &             & = f_s(T) \rho_s(T) + f_n(T) \rho_n(T), \end{eqnarray}
where, the average densities of the respective fluids are
\begin{eqnarray}  \rho_t(T) = {N_t \over V_t} =(1/V_t)\int_{V_t}  \rho_t(\rb, T) d\rb,   \quad \quad  \  t=s, n.  \end{eqnarray}

\begin{figure} [here] \begin{center} \leavevmode \hbox{%
\epsfxsize= 9 cm     \epsfbox{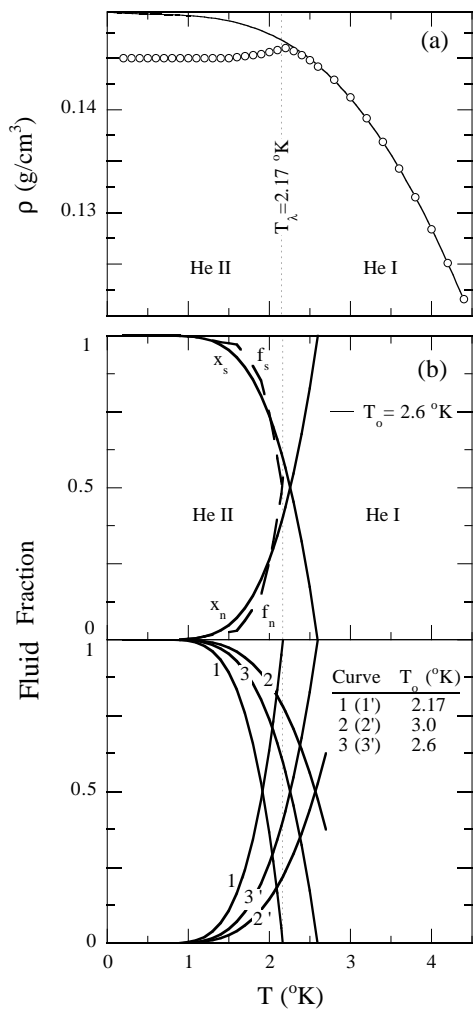}} \end{center} 
\caption{  (a). The density $\rho_n(T)$ of the normal fluid component of He II, solid line within $T<T_{\lambda}$, as is extrapolated using the thermal expansion function Eq. (\ref{eq-5b}).  The coefficients are determined by a least squares fit to the experimental data of $\rho_n(T)$ (circles)  \ref{kerr:el} between 2.6 - 4.4 \oK. \ (b). The volume fractions  $f_s(T)$ and  $f_n(T)$ of the superfluid and the normal fluid of He II, dashed curves, as given by Eq. (\ref{eq-5c}), Sec. \ref{Secz4}, established based on the composite structure of He II, namely large separate aggregates of the two fluids, and with the experimental density of He II as input.  The number fractions $x_s$ and $x_n$ of the superfluid and the normal fluid of He II, solid lines, obtained from thermodynamic evaluation using Eqs. (\ref{eq3-28})-(\ref{eq3-28})$'$, Sec. \ref{Secz9}. The lower and upper limits, solid lines 1 and 2, are obtained using $T_o=2.17$ and 3.0 \oK, \ and  the optimum values, solid lines in the upper graph and solid lines-3 in the lower graph of (b), are obtained using $T_o=2.6$ \oK. 
\label{fig-frac} }   \end{figure}

Based on Eqs. (\ref{eq2})-(\ref{eq4}) we now 
determine $f_s(T)$ and $ f_n(T)$ using only the experimental densities of liquid $^4$He \cite{kerr:el} as input data. Based on the close resemblance of the superfluid to a solid in a few important ways, e.g. as is shown by the essentially $T$ independent thermodynamic functions (Sec. \ref{Secz9}), we can assume that $\rho_s$ is basically independent of $T$ on the scale of concern, and can then write $\rho_s \approx \rho(0)$, $\rho(0)$  ($= 0.1450 \quad \rm{g/cm}^3$) being the He II density at zero Kelvin.  To obtain the $\rho_n$ of He II, we next extrapolate $\rho_n(T)$ of the He I phase into the He II phase region  using the thermal expansion function:
\begin{eqnarray} \label{eq-5b} 
\rho_n(T) = {0.1495 \ov \left(1 + 0.001603T - 0.002035T^2 + 0.001219 T^3\right)^3}. \end{eqnarray}
Where, the numerical coefficients are obtained from a least squares fit (solid line, FIG. \ref{fig-frac}a) to the experimental $\rho(T)$ data of He I (circles) between 2.6 to 4.4 \oK. \   Substituting into Eq. (\ref{eq4}) with the experimental  $\rho(T)$ 
and the extrapolated $\rho_n(T)$ given in Eq. (\ref{eq-5b}) for $T<2.17$ \oK, \  and by further combining with Eqs. (\ref{eq2})-(\ref{eq3}), we then obtain for He II
\refstepcounter{equation}
$$\displaylines{ \label{eq-5c}      f_s(T)  = \left(\rho(T) -\rho_n(T)\right) / \left(\rho_s - \rho_n(T)\right)   \hfill  (\ref{eq-5c}a)  \cr
= { \rho(T) - {0.1495 \ov \left(1 + 0.001603T - 0.002035T^2 + 0.001219 T^3\right)^3 }
\over  0.145 - {0.1495 \ov  \left(1 + 0.001603T - 0.002035T^2 + 0.001219 T^3\right)^3}}      \cr
{\rm and} \quad f_n(T) = 1- f_s(T),   \hfill      (\ref{eq-5c}b)      }$$  
\input epsf  
\noindent
 where all densities are in \rm{g/cm$^3$}. 
The resulting $f_s(T)$ and $f_n(T)$ (dashed lines in FIG. \ref{fig-frac}b) vary rapidly with $T$ from about 50 \% to 100 \%  and conversely between 2.17 - 0.6 \oK, \ and approach to constant below 0.6 \oK. \ In comparison, by Eq. (\ref{eq-5b}) $\rho_n(T)$ and $\rho_s(T)$ vary by less than 4 \%  across the entire $T$ range of He II. Combining the above features with  (\ref{eq4}), we therefore obtain that the $T$-dependence of $\rho(T)$ throughout the transition in 0.6-2.17 \oK \  is principally determined by the variations in $f_s(T)$ and $ f_n(T)$ with $T$.  It is further particularly noticeable that at $T_{\lambda}$, $f_s$ ($\sim 0.5$) and $f_n$ ($\sim 0.5$) are not zero and one. The approximations involved would not cause any significant error as has been tested; rather, the non-unity of the fractions at the $T_{\lambda}$ is plausibly as a consequence of the second order phase transition. For, in such a transition a measurable $n$ to $s$ conversion begins to occur at a $T$, denoted by  $T_o$, well above \Tlmb; \ that is, $f_s(T_o) =0$ and $f_n(T_o)=1$. The physics involved will be further clarified in Sec. \ref{Secz10}. The $f_s(T)$ and $f_n(T)$ obtained above are in good agreement with the thermodynamic results for $x_s, x_n$ in Sec. \ref{Secz9}, the resulting specific heat and entropy obtained based on which are in good agreement with experimental data;  and these are consistent with the characteristic variation across  the He II temperatures exhibited by a variety of properties of He II, including specific heat, and second sound.

\section{ Equations of motion, the solutions, and their use for directly predicting properties of He II } \label{Sec5} \label{Secz5}

\subsection{The equations of motion via coordinate decomposition } \label{Sec5.1} \label{Secz5.1}
The experimental evidence discussed in the preceding sections has pointed to the separate manifestations of the superfluid atomic dynamics with differing time and energy scales.  Namely, a momenton oscillation, being of the scale of $\sim 10^{3}$ cycles/sec, is negligibly slower compared to the single atomic oscillation with respect to $\xib$, typically of the scale of $10^{11}$ cycles/sec, so that the former can be regarded as being stationary within the oscillation cycle of the latter. Furthermore, the diffusion rate of the helium II atoms, with a diffusion constant $\sim 10^{-8}$ cm$^2$/sec. (cf. Sec. \ref{Sec5.4}), 
is negligibly slower compared to the single atom oscillation rate, and will on the other hand neither effect the momenton oscillation.  
We now establish the equations of motions accordingly via decomposing the total motion in a similar way to the above. First, we may express the absolute position $\rb $ of an atom with the vector sum
\begin{eqnarray} \label{eq5}
&	\rb (x,y,z)  = &\ub(u_x,u_y,u_z) + \Rb(X,Y,Z) \\ \nnr  & & + \xib(\xi_x,\xi_y,\xi_z) + \Xib(\Xi_x,\Xi_y,\Xi_z),  
\end{eqnarray}
where $\ub$ is the displacement of a He atom relative to $\Rb$, $\Rb$ is the equilibrium position of the atom with respect to $\xib$, $\xib$ is the displacement of the center-of-mass of a momenton relative to $\Xib$, and $\Xib$ is the equilibrium position of the momenton with respect to the laboratory system 
(FIG. \ref{fig-mom}b).
Accordingly, the total interatomic potential is $V(\Rb+\ub)+\V(\Xib +\xib)$, with the 1st and 2nd terms being due to the atoms, as viewed from $\xib$, and the momentons. As discussed above, the relative coordinates are effectively decoupled, meaning $\pd \rb / \pd x = \pd \ub / \pd x + \pd \Rb / \pd x + \pd \xib / \pd x + \pd \Xib / \pd x$, 
etc. Hence, we obtain four separate equations of motions for an atom at site $l$ of mass $m$ and an momenton $j$ of mass $M_j$, which in the semi-classical form formally are the Newton equations: 
\refstepcounter{equation}
$$\displaylines{  
\label{eq7} \hfill {\bf F}_l =  m {\pd ^2 \ub_{l}(K,t) \over  \pd t^2} \hfill  (\ref{eq7}a)          \cr 
\refstepcounter{equation}  \label{eq8} 
 \hfill  {\bf F}_R = m {\partial^2 \Rb_{l}(K_b, t) \over \partial t^2}, \hfill  (\ref{eq8}a)  \cr
\hfill \qquad\qquad\qquad\qquad l=1,2, \ldots   \hfill  \cr   
\refstepcounter{equation}  \label{eq9} \hfill {\bf F}_{\xi}=  M_{j} {\pd ^2 \xib_{j}(K_M,t) \over  \pd t^2}, \hfill   (\ref{eq9}a)  \cr
\refstepcounter{equation}    \label{eq10}
\hfill  {\bf F}_{\Xib}=  M_j {\pd ^2 \Xib (k_s, t) \over  \pd t^2}, \hfill  (\ref{eq10}a)  \cr
\hfill \qquad\qquad\qquad\qquad j=1,2, \ldots. \hfill  
}$$
Quantum mechanically the total Hamiltonian and the wavefunction are   
\begin{eqnarray}\label{eq-H} 
 H= H_u + H_R + H_{\xi}+ H_{\Xi},  \\ \label{eq-psi}  
 \psi(\rb) = \varphi(\ub) \phi(\Rb) {\mit \Phi}_{\xi}(\xib) {\mit \Psi}_{\Xi}(\Xib).
\end{eqnarray}
Here $H_u, H_R, H_{\xi}$ and $H_{\Xi}$, and $\varphi, \phi, \Psi_{\xi}$ and $\Psi_{\Xi}$ are the Hamiltonians  and wavefunctions in the respective coordinates, $\ub, \Rb, \xib$ and $\Xib$. The counterparts of Eqs. (\ref{eq7}) - (\ref{eq10}) are the Schr\"odinger equations:  
$$\displaylines{ 
\hfill H_{u}  \varphi  = \e(K) \varphi          \hfill   (\ref{eq7}b) \cr 
\hfill  H_{R} \phi =  \eR  (K) \phi             \hfill  (\ref{eq8}b)  \cr
\hfill  H_{\xi}{\mit \Phi}_{\xi} = \in (K_M){\mit \Phi}_{\xi}     \hfill      (\ref{eq9}b) \cr 
\hfill H_{\Xi} {\mit \Phi}_{\Xi}(\Xib) = e_s(k_s) {\mit \Phi}_{\Xi}(\Xib). \hfill  (\ref{eq10}b) 
}$$

In Secs. \ref{Sec5.2}-\ref{Sec5.5} we set up explicitly for the above equations and solve, and based on the solutions predict a few immediate properties of the system.

\subsection{ Cooperative oscillation of single atoms and single phonon excitation } \label{Sec5.2} \label{Secz5.2}

\subsubsection{Set-up of Eqs. (\ref{eq7}a)-(\ref{eq7}b)}  \label{Secz5.2.1} 
Consider in the $\xib$ coordinates the superfluid atoms are, as discussed in Sec. \ref{Secz2.1}, localized and each oscillates about its equilibrium site $\Rb (X,Y,Z)$ along a chain of atoms say in the $X$ direction, in a potential well $V(X)$ (FIG. \ref{fig-harmpot2}a). The mean separation of the atoms is $a$; as based on the  $q_b$ (Sec. \ref{Secz2}), $a$ $=2\pi/q_b=$3.3 \AA. \ 
The complete oscillation motion can be projected along three orthogonal chains in $X,Y, Z$ directions, which can be oriented in an arbitrary direction as the fluid is isotropic. In three dimension the dynamics above can be accordingly represented for a simple cubic {\it pseudo} lattice (see also Sec. \ref{Secz2}), which we shall use below, of an apparent and effective lattice constants, $a$ and  $a_r$ (to be explained below), respectively, with one atom per unit cell.
When needed, the effect of fluctuation about $a$ may be plugged in e.g. by damping the average result obtained this way.  As the fluid has no shear elasticity, the different parallel atomic chains are not correlated in motion or coordination; and along each chain only longitudinal waves present. 

 The exact many-body potential energy of an atom at $X_l$ is $V(X_l+u_l)$, which is a function of the displacements of all ($\Nu$) atoms within a correlation length $\Up$: 
\begin{eqnarray}\label{eq-Vx} V(X_l+u_l(\uls)) = V (X_l+u_l(\uls); \uons, \u2s,  \ldots, \uNs) \end{eqnarray}    
where $\uls=\sum_{l'} \lf[ (X_l+u_l)- (X_{l'} + u_{l'})\rt]$ is the displacement of the atom at $X_l$ relative to all other atoms in the chain. $V(X_l+u_l) $ expands into the Taylor series: 
\begin{eqnarray}V(X_l+u_l) =  V_0 +  {d V \ov d\uls} \uls + {1\over 2} {d^2 V \ov d\uls^2} \uls^2 + \ldots.  \end{eqnarray} 
where $ V_0$  is just the semi-classical ground-state potential energy with a zero-point energy  effectively included; $ V_0$ will not contribute to the force except when the motion of $X_l$ is in question in which case it together with the $X_l(t)$ motion is then separately represented in Eq. (\ref{eq8}a) In the LOD scheme, $ {d V \ov d\uls} \uls$ will average to zero for the atom oscillating about its equilibrium site. Because the superfluid bonding (Sec. \ref{Secz3}) has a large spatial extension, we expect that at  larger $u_l$ values, ${d^2 V \ov d\uls^2}$ varies only slowly with $u_l$, hence higher order terms than $d^2 V / d\uls^2$ are small altogether and can be neglected even at a relatively large $u_l$.  $V(X_l+u_l)$ then reduces to the quadratic
\begin{eqnarray}\label{eq:pot:1}  V(u_l) = V_0 + {1\over 2} {d^2 V  \ov d\uls^2} \uls^2
=  V_0 + {1\over 2} \aon \uls^2;\end{eqnarray}
where $X_l$, being inexplicit, is dropped for simplicity, $u_l=u_l(\uls)$ and $\aon ={d^2  V \ov d\uls^2}$. For the many-body representation (\ref{eq-Vx}): 
\begin{eqnarray}\label{eq-alf} \aon = {d^2 V  \ov d\uls^2}=\sum_{l'}{\pd \ov \pd \ulsp} \lf({d V \ov d \uls}\rt)   \nonumber \\
 = \sum_{l'} {\pd^2 V \ov \pd \ulsp \pd \uls} {\pd \ulsp \ov \pd \uls}, \end{eqnarray}
where ${d V \ov d\uls}= \sum_{l'} {\pd V \ov \pd \ulsp} = {\pd V \ov \pd \uls}$ with all  other terms cancelled out as the displacements altogether are symmetric about $X_l$. 
Taking derivative with respect to $u_l$ on both sides of (\ref{eq:pot:1}) we get the force acting on the atom at $X_l$: 
\begin{eqnarray} \label{eq-force}
F_l = {d V \ov d \uls}= \sum_{l'} {\pd^2 V \ov \pd \ulsp \pd \uls} {\pd \ulsp \ov \pd \uls} 
\simeq  - \aon \uls.  \end{eqnarray} 
Eq. (\ref{eq-force}) or (\ref{eq-alf}) contains only the non-zero, diagonal terms $q=q'$ ($q, q'=X,Y,$ or $Z$) of the block matrix components ${\pd^2 V \over \pd \ulsp^{q'} \pd \uls^q}$,  
representing the absence of shear elasticity of the fluid. Further we sum for $\uls$ over only nearest-neighbors:  $\uls =-u_{l+1}-u_{l-1}+2u_{l}$. 
This together with the last expression of (\ref{eq-force}) represents an effective nearest-neighbor interaction and is used below; and the many-body correlation effect will be effectively included in an effective  $\aon$ (or $c_1$) value determined by fitting to experimental data. With the $\uls$ above in (\ref{eq-force}), and in turn (\ref{eq-force}) for $F_l$ in Eq. \ref{eq7}a  we get:
$$-\alpha_1 (-u_{l+1}-u_{l-1}+2u_{l})  = m {\partial^2 u_l \over \partial t^2},  \qquad l=0,1,\ldots \eqno(\ref{eq7}a)'$$
this being the independent harmonic oscillator description for the cooperative motion of the superfluid atoms in the $\xib$ coordinates. It is noteworthy that, first, the averaging over many neighboring atoms in Eq. (\ref{eq-alf}) results in an identical $\alpha_1$ for all sites at all time, despite the reality that the physical parameters at each site in general fluctuate. This thereby leads to the well-defined, site-independent excitation energy as from Eq. (\ref{eq7}a)$'$ (similarly Eq. (\ref{eq7}b)).
Second, $F_l$ is a restoring force, either attraction or repulsion, as long as sign$(F_l) = -$ sign $(\uls)$, 
and further for the given $|F_{l}| \propto |\uls| $ here, $V(u_l)= -\int F_l d\uls$ will then yield Eq. (\ref{eq:pot:1});
we shall encounter this notion below.

From the above and using in advance the $\uls $ expression from solving Eq. (\ref{eq7}a) in Sec. \ref{Secz5.2}c, we have  $V-V_0= {{\aon} \ov 2}\uls^2={1\ov 2}m \w^2 u_l^2$, where ${\aon}(K)=m\w^2(K)$, and can then express for the total energy: $U_K= {p_l^2\ov 2m} + (V-V_0)$ for the normal modes of each oscillator. The $U_K$, with $p_l= -i\hbar {\pd^2 \ov \pd u_l^2}$, directly leads to  $H_u$, with which 
in Eq. (\ref{eq7}b) we have: 
$$  \left( - {\hbar^2 \ov 2 m } {\pd^2 \ov \pd u_l^2} + {m \w^2  \ov 2} u_l^2   \right)  \varphi(\ub_l)     = \e(K) \varphi(\ub_K) \eqno(\ref{eq7}b)'$$

\subsubsection{ Solution of Eq. (\ref{eq7}b), single phonons and their wavefunctions } \label{Secz5.2.2}
Equation (\ref{eq7}b)$'$, which we here discuss first,  has the well-known eigen energy solution:
\begin{eqnarray} \label{eq2-8}
\epsilon_n (K) = \left(n + {1 \over 2} \right) \hbar \w(K) \quad \quad n=0, 1, 2, \ldots,  \end{eqnarray}
which consists of $n$ phonons, 
each having an energy quantum 
\begin{eqnarray} \label{eq-EK} \E(K) = \hbar \omega (K). \end{eqnarray} 
The stationary eigen functions are (here $u\equiv u_l$)
\begin{eqnarray}\label{eq-vph} 
\varphi_n(u) = c_n H_n\left(\sqrt{{m\omega  \over \hbar}} \ u\right) \exp(- {m\omega \over 2 \hbar } u^2). \end{eqnarray}
$H_n $ are the Hermit polynomials and $c_n$ the normalization coefficients.
 The expectation value $<\epsilon_n> = \int \Phi_n^*(\ub, t) \epsilon_n \Phi_n (\ub, t) d \ub = \sum_n \sum_K \e_{n}(K)$, where $\Phi_n(u, t)=\varphi (u) \exp(\imath H/\hbar t)$ are the time dependent eigen functions, yields the total energy of the system, which  will be given in Sec. \ref{Secz9} by an ensemble average, the internal energy $U_s-U_{s0}$.

Equation (\ref{eq-vph}) informs:
1) The spatial extension of $\varphi(u)$, as can be measured by  $\La_u = 2u|_{\varphi(u) =0.5\varphi(0) }= 2  \sqrt{{2\hbar\over m\omega} \ln 2}$, is large for the small helium mass. At a low energy, e.g. $\omega = 1 $ \oK, \ $\La_{u_{He}} \simeq 8.2 $ \AA \ and is a few times larger than $a$. The minimum-energy (of the $\ub$ motion) wavefunction $\varphi_0(u) = \sqrt{{m\omega \over 2\pi\hbar}} e^{{m\omega \over 2 \hbar } u^2}$ at this $\w$ is seen to extend into and beyond the first nearest-neighbor (solid curves, FIG. \ref{fig-harmpot2}a2); and similarly with the excited states. By contrast, the wavefunction for an argon atom, of a mass $m_{{\rm Ar}} \approx 10 m_{{\rm He}}$, (dotted curves in the figure) has $\La_{u_{Ar}} \simeq 2.6$ \AA, \ which is less than the interatomic spacing $a_{{\rm Ar}} \simeq 5.3$ \AA \ taken for solid argon. 
$\La_u$, being a similar measure to the thermal de Broglie wavelength,
thus defines that, (viewed now in $\xib$) a helium is quantum mechanical and an argon classical.  $\La_u$ increases with $\w$, but by Sec. \ref{Secz5.2.2} at a larger $\w$ the atom is then confined in a narrower potential well; and at a high $\w$ comparable to $\Db$ the atom converts to the state of  a normal fluid atom.  
2) It follows from the large $\La_u$ that the single particle momentum of helium cannot be determined better than $ \hbar / \La_u$.
3) The immediate consequence of the large $\La_u$, which causes the wavefunctions of several nearest-neighbor atoms or so to be entangled, is to result in the many-quantum-atom correlation scheme, and the consequent substantial superfluid bond and the localization of the He atoms, as discussed in Sec. \ref{Secz3}.
4) However, $\La_u$ is only several times $a$ or so say within a region $\Up$; hence the helium atoms are "indistinguishable" effectively within only the limited  $\Up$ region  ($\Up<$ or $<<$ $V$ of the container), in which the atomic wavefunctions overlap. This is in contrast to a free quantum particle whose uncertainty region consists in the entire volume of the particle assembly (the free electron gas is an example of such). 
Furthermore, since the atoms are localized and each can most probably be found about its equilibrium site, thus the individual atoms are not "identical" in respect of the sites they can occupy. 
Restating this condition, by regarding the localized atoms as being effectively "identical" instead, we have that  the centres of mass of the atomic wave packets are effectively "distinguishable"; the atoms will become "effectively  identical" if one views each "atom wavefunction + site"; the "independent harmonic oscillator" representation in this section is in effect a formal transformation to this status. Compare the discussion relevant to the $R$ motion in Sec. \ref{Secz5.4}. 

\subsubsection{ Solution of Eq. (\ref{eq7}a), the dispersion curve for $q<1.93$ \AAi, \ and the potential energy} \label{Secz5.2.3} 
\begin{figure} [here] 
\begin{center} \leavevmode \hbox{%
\epsfxsize= 6.5 cm     \epsfbox{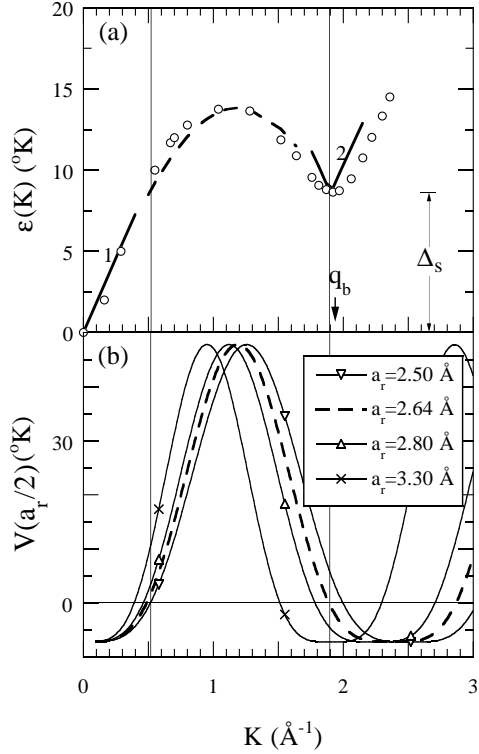}} \end{center} 
 \caption{  
(a) Theoretical total excitation energy $\ev(K)$ versus phonon wavevector $K$ of the superfluid of He II, solid and dashed curves. The excitation at a $K<K_b-\sg/2$ is due to the creation of a single phonon, $\E(K)$ (solid curve 1), as predicted from the solution (\ref{eq2-7b}a-b) of the equation of motion of the He atoms performing independent harmonic oscillations relative to $\xib$ in the potential wells $V(u)$ shown in FIG. \ref{fig-harmpot2}. The above $K$ region further divides into $K \alt 0.6$ \AAi (Eq. \ref{eq2-7b}a) and  $<0.6K <K_b-\sg/2$ \AAi \ (Eq. \ref{eq2-7b}b), where the apparent  parameter $a=3.3$ \AA \ and the effective $a_r=2.64$ \AA  \ are, respectively, used as the interatomic spacing. 
In the vicinity of $K_b=1.93$ \AAi, \ i.e. $|K-K_b| <\sg/2$, $\ev(K)$  (solid curve 2) primarily results from the excitation of a superfluid bond, of an activation energy $\Ds$, superposed with a small single phonon excitation energy: $\ev(K)=\Ds(K) + \E(K)$, where $\ev(K_b) = \Ds$.   Neutron diffraction data points at 1.12 $\rm{^oK}$\cite{henshaw:w:1961} are represented by circles.
(b) Potential enenrgy of a superfluid atom, $V(u)$, as a function of $K$, Eq. (\ref{eq:pot:1})$'$ with $u=a_r/2$, plotted for several $a_r$ values from 2.5 to 3.3 \AA. At a fixed $a_r$, $V({1\ov 2}a_r)$  has a positive peak in a certain $K$ region which shifts towards the higher $K$ end with decreasing $a_r$.  Only if $a_r \approx 2.64$ \AA, \ the $K$ region where $V({1\ov 2}a_r)>0$ (dashed curve) coincides with the $K$ region where $\E(K)> \Delta_s$ (dashed curve, FIG. \ref{fig-wVK}a); this $K$ region is bounded between  (0.51, 1.87) \AAi \ as indicated by the two dotted vertical lines. \label{fig-wVK} 
 } \end{figure}

 Equation (\ref{eq7}a) can be readily solved to give the oscillation amplitude $u_l(K) = A e^{-\imath( l a_{\a} K + \w t)}$, 
thus $\uls = 4u_l \sin^2 ({Ka_{\a}\ov 2})$, and the normal-mode angular frequency and the $\E(K)$ by  combining with (\ref{eq2-7b})  (both in J) 
\begin{eqnarray} \label{eq2-7b}
& &  \w(K) 
= \sqrt{{\alpha_1 \ov m}} \quad  2 \sin \left({K a_{\a} \ov  2} \right),  \nonumber  \\
& & {\rm and} \quad \ev(K)= \E(K) = \hbar \w(K)  \nonumber  \\ 
& &  {\rm where:}   \nonumber  \\ 
& & a_{\a}= \left\{
\begin{array}{ccc}
\ \ a,   \qquad  \ K< 0.51, \ {\rm or }  \ 1.87 <K <K_b -{\sigma\ov 2} \AA^{-1}  \nonumber  \\
\qquad\qquad\qquad\qquad\qquad\qquad (\ref{eq2-7b}a)      \nonumber  \\
\ a_r,  \qquad     \qquad   \  0.51 \le K \le 1.87 \  \AA^{-1}  \quad  \quad (\ref{eq2-7b}b) \end{array} \right.  \nonumber
                                           \end{eqnarray}
For $K\alt 0.5$ \AAi \ ($Ka << 1$), (\ref{eq2-7b}a) becomes
$$  \E(K) \simeq \hbar c_1 K. \eqno(\ref{eq2-7b}a)^{\prime}$$
 The phase velocity here is $\E/(\hbar K) = c_1 = \sqrt{\alpha_1/m} \ a$, 
being equal to the propagation velocity of the sound wave, $(1/\hbar)(\partial \E/\partial K)$. Parameterizing  Eq. (\ref{eq2-7b}a) using the pulse transmission value $239$ m/sec \cite{atkins:stasior} for $c_1$, we get $\E(K)= 18.25 K$ (\oK) (solid line-1, FIG. \ref{fig-wVK}a and also FIG. \ref{fig-ex-spec}b) where $K$ in \AA$^{-1}$, in satisfactory agreement with the neutron-probed excitation curve \cite{henshaw:w:1961} (circles) in this $K$ region.  
The slope of the latter from a least squares fit is $c_1 = 238.9$ m/sec. The above result has been used as a fed-back information in Secs. \ref{Sec2.1}-\ref{Sec2.3}.

\input epsf  

\begin{figure} [here] \begin{center} \leavevmode \hbox{%
\epsfxsize= 8.5 cm     \epsfbox{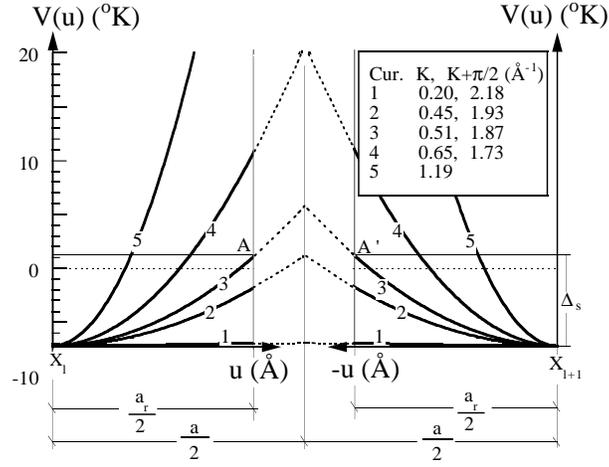}} \end{center} 
\caption{ Potential energy $V(u)$ versus $u$ a of superfluid atom centered at $X_l$ or $X_{l+1}$, Eq. (\ref{eq:pot:1})$'$ with $a_r=2.64$ \AA, \  plotted for several $K$ values as indicated in the inset.  
For $K=0.51$ \AAi,\ from which towards higher $K$ values $\ev(K)>\Ds$ (see FIG. \ref{fig-wVK}), the potential turns to be repulsive, i.e. $V(u)>0$, at the effective border $u \pm {a_r \ov 2}$ (curve 3). $a$ is the apparent interatomic distance of He II as given by the reciprocal of $K_b$: $a=2\pi/K_b=3.3$ \AA. \label{fig-Vu}  } \end{figure}

For 0.5 $\alt K\alt K_b (\equiv q_b)$ \AA$^{-1}$, $\E(K)$ involves explicitly $a_{\a}$ and here $a_{\a}=a_r$. $a_r$, as an effective quantity of $a$, was introduced when setting up Eq. (\ref{eq7}a)$'$ in order to now account for an apparent distortion of $\E(K)$ from the experimental curve if $a=3.3$ \AA \ is used in Eq. (\ref{eq2-7b}). For we recognize that, a helium atomic oscillator, having a broad wavefunction as Sec. \ref{Secz5.2}b showed, would overlap with its neighbours and hence repel each other at an effective border at $|{1\ov 2}a_r|< |{1 \ov 2}a|$ (FIG.\ref{fig-Vu}). If the excitations indeed include oscillations of amplitudes larger than ${1\ov 2}a_r$, 
then the finite oscillator size, reflected by the repulsion length $AA'$ in FIG. \ref{fig-Vu}, must be excluded from the apparent $a$. Thereby at these modes the atomic oscillator is effectively confined in $(-{1\ov 2}a_r, {1\ov 2}a_r)$. In resulting in a $\E(K)$ (dashed line in FIG. \ref{fig-wVK}a and also FIG. \ref{fig-ex-spec}b) in agreement with the experimental data in this $K$ region, we find $a_r \approx 0.8 a =2.64$ \AA. \  
We will shortly justify that with this $a_r$ value, $V(u)$ indeed turns to be $>0$ at $u= {1\ov 2}a_r$. 
\input epsf  

Using in Eq. (\ref{eq:pot:1})$'$ the $\alpha_1 =  {c_1^2 m\ov {a_r}^2}$ as determined from the linear $K$ region relation and the $\uls$ expression, both just given, 
and with $ V_0 \simeq U_{0s}= -7.2$ \oK/atom  as determined from experiment \cite{simon:el:1950}$^,$\cite{london:1954} (note that by letting $V_0\simeq U_{0s}$ the  zero point energy is effectively included; see also Secs. \ref{Sec2.3}, \ref{Secz9}) we have
 $$ V(u_{l}) = V_0+ {1\ov 2}{c_1^2 m \ov  {a_r}^2} \uls^2   
= -7.2 + 31.6 u_l^2 \sin^4({Ka_r \over 2}),
\eqno(\ref{eq:pot:1})' $$  
where  $V(u_{l}) $ is in \oK, \ $u_l$ in \AA, \  and $K$ in  \AAi; \ 
  $a_r=2.64$ \AA \ for the parameterization; the resulting $V(u)$ for selected $K$ values are graphically shown in FIG. \ref{fig-Vu}. For $K < 0.51$ or $K> 1.87$ \AA$^{-1}$, \   $V(u)<0$ for all $|u|$ values up to ${1\ov 2}a_r$. 
For $K=0.51$ (or symmetrically 1.87) or  $0.51 < K < 1.87$ \AAi, $V(u)$ turns to be repulsive at $u ={1\ov 2}a_r$ or earlier  -- we preassumed this in the preceding paragraph. 
Furthermore, the repulsive feature of $V(u)$ with $K$ in $(0.51, 1.87)$ \AAi \ is also precisely as expected from the excitation scheme derived in Sec. \ref{Sec2}. Namely, since in this $K$ region the excitation energy $\E(K)>\Delta_s$ and $\E(K)$ corresponds to the $\max(|u|)$ for the given $K$, the $\E(K)$ here must therefore result from repulsion. 

Compute $V(u_l)$ as a function of $K$ with $u_l={1\ov 2}a_r$ in Eq. (\ref{eq:pot:1})$'$, for different $a_r$ values from 2.5 to 3.3 \AA. \  We find that at a fixed $a_r$, $V({1\ov 2}a_r)$  has a positive peak in a certain $K$ region which shifts towards the higher $K$ end with decreasing $a_r$ (FIG. \ref{fig-wVK}b). Only if $a_r \approx 2.64$ \AA, \ the $K$ region where $V({1\ov 2}a_r)>0$ (dashed curve) coincides with the $K$ region where $\E(K)> \Delta_s$ (dashed curve, FIG. \ref{fig-wVK}a). 
Thus we have verified that the solution $a_r \approx 2.64$ \AA \ is unique.

Furthermore we note that at $K=K_b =1.93$ \AAi, or $K a =2\pi$, all atoms move in phase: $u_{l+1}/u_l ={ \exp(l 2\pi+\w t)\ov \exp((l+1)2\pi+\w t)} \equiv 1$, 
and $\uls \equiv 0.$  That is, in the $K_b$ mode, in real space all atoms are at the bottoms of their potential wells as viewed, for example, by a neutron scattered due to this mode. The oscillations at the other normal modes would cause, with respect to the neutron, a snapshot of somewhat irregularly displaced "equilibrium sites" of the atoms which will add to their already irregularity due to a short range ordering.

Finally, the character of the localized oscillation of the helium atoms, being clearly distinct from that in solids, has up to now been represented in the formal treatments and are re-emphases as follows: it is relative to $\xib$ rather than $\Xib$, it is in cages fluctuating in location and dimension, and it is in a shallow potential well which switches to repulsion at larger oscillation amplitudes.

\subsection{ Single atom translation and excitation of the superfluid bond  } \label{Sec5.4}\label{Secz5.4}
\subsubsection{The semi-classical representation Eq. (\ref{eq8}a), excitation of the superfluid bond} \label{Secz5.4.1}
Equation (\ref{eq8}a) describes, in a superfluid equilibrium in the absence of a pressure and composition gradient, the thermal fluctuation resultant translation (self-diffusion) of the equilibrium position $\Rb$ of a helium atom.  This motion can be tracked by solving the simultaneous equations of motion for $N$ atoms with an interatomic potential $V_0$, e.g. numerically with the help of computer. 
 Macroscopically, the motion of, say, $N'$ tagged atoms form a thermal diffusion flux of a concentration  $c(\Rb)$ ; the diffusion equation e.g. along $X$ axis is 
 $$ D \divd^2 c(\Rb,t) - {\pd c(\Rb,t) \over \pd t} = 0 \quad  {\rm (the \ macroscopic \ form)}. \eqno(\ref{eq8}a)' $$
 $D$, the diffusion constant, can be estimated by 
\begin{eqnarray} \label{eq-dif} & D  & = C a^2 {\w \ov 2 \pi} e^{-\Delta_s k_BT} \nonumber \\
& & \approx 2.4 \times 10^{-8} \qquad {\rm cm}^2 {\rm/sec}
 \end{eqnarray}
where  $C$ is the coordination number, $e^{-\Delta_s/k_BT}$ the Boltzmann factor and, by Sec. \ref{Secz2.2}, $\Delta_s$, $=$8.6 \oK, \ is the activation energy of a superfluid bond.
The numerical value of $D$ is obtained using the previous values for $a$, and $\omega (=10^{-9}$ 1/sec), with ${C\ov 2\pi}\approx 1$ and $T=$ 1 \oK. \
This $D$ value is as low as that of diffusion in a solid, informing that a superfluid atom is virtually localized.
Accordingly the solutions of (\ref{eq8}a) and (\ref{eq8}a)$'$, of which $c(\rb)$ is known to be a Gaussian function, are not of interest here.  $D$ represents the probability for an helium atom to translate from its bonding state at $\Rb$ to that at a neighboring site at $\Rb'$ (which can statistically also be $\Rb$), by overcoming an energy barrier height $\Delta_b + \dv$ (FIG. \ref{fig-harmpot2}a):
$$ P(\Rb \rightarrow \Rb') \propto D \propto  e^{- {\Delta_{s\alpha} \ov  k_B T}}.$$
With He II, the $P(\Rb \rightarrow \Rb')$ processes of practical interest are ones caused by external perturbations, and in particular, the manifestations of the processes in the excitation spectrum which we discuss below, and in the specific heat which we discuss in Sec. \ref{Secz9}.

Upon a local perturbation the excitation at $K=K_b $ is by Sec. \ref{Secz2.2}(ii) due to activation of the superfluid bond: 
$$ \label{eq-ex1}  \ev (K) = \Delta_s = 8.56 \qquad (^o{\rm K})        \qquad          K=K_b   \eqno(\ref{eq2-7b}c)   $$
And at a $K$ apart from $K_b$ but in a narrow vicinity $ 0<|K-K_b| \le \sigma$ is, by combining with Secs. \ref{Secz2.1} and \ref{Secz5.2}, therefore
$$ \displaylines{
\hfill  \ev (K)  = \Delta_s(K_b) +\E(K)    
\approx  8.56 + 18.18  |K-1.93|  \quad (^o {\rm K}), \hfill                 \cr
\hfill  0<|K-K_b| \le {\sigma \ov 2}   \qquad     (\ref{eq2-7b}d) 
}$$
Here, $\E(K)= \hbar c_1  |K-K_b| $ is the single phonon excitation energy of Eq. (\ref{eq2-7b}) after the simplification: 
$ - \sin(\pi - {Ka\ov 2}) 
\simeq - (\pi - {Ka\ov 2})$ for $\pi - {Ka \ov 2} <<1$. 
 For the parameterization: $\Delta_s = \Delta_b + \dv = 8.6$ \oK \ and $c_1 \simeq 239$ m/sec as given previously; $K$ is in \AA$^{-1}$. \  The resultant $\ev(K)$ (solid line-2 in FIG. \ref{fig-wVK}a and also FIG. \ref{fig-ex-spec}b) agrees fairly well with the experimental curve, except for its hard turn at $K_b$ as contrasted to the experimental smooth turn.
This differnece is as expected, since the theoretical treatment does not include the effect of the fluctuation (resulting from a short range ordering) in the superfluid bond length  which occurs in the real superfluid. 
Indeed, experimntal excitation data at high pressure ($\sim$25 atm) \cite{Henshaw:Woods:1961}, which causes a "solidification" effect,  has shown a sharp turn at $K_b$ similar to the theoretical description.
(There are also shifts of the dispersion curve to a higher frequency and to a larger period, which can be understood similarly.)  

As $K$ departs from $(K_b-{\sigma\ov 2},K_b+{\sigma\ov 2})$ to lower $K$ values,
the probability of the $\Delta_b$ excitation will drop rapidly, for the liquid density $n(\rb)$ drops as  $\rb$ departs from $\Rb$. So that $\ev(K)=\E(K) $, which is the purely phonon excitation energy  as of Eqs. (\ref{eq2-7b}a)-(\ref{eq2-7b}b).

\subsubsection{The quantum mechanical representation Eq. (\ref{eq8}b)} \label{Secz5.4.2}
The potential energy for the $R$  motion for a reference atom at $R$ is:   
\begin{eqnarray}   \label{eq-Pp} 
  \VRo  (\le 0)= \left\{ \begin{array}{cc}   
     V_0  -  \Do,    \qquad   |X|\le {a\ov 2}  \cr
    0,          \qquad\qquad\quad  |X|>0,
\end{array} \right. \end{eqnarray} 
Here $\Do$ is  the zero point kinetic energy of the $R$ motion; $V_0=-7.2 $ \oK/atom as before. 
With $H_R= - {\hbar^2 \ov 2 m} \nabla^2 + \VRo$, $\VRo$ as expressed above, in  (\ref{eq8}b) we have in the $X$ direction
$$\displaylines{ \hfill  
\lf(- {\hbar^2 \ov 2 m} {d^2 \ov X^2} +  \VRo    \rt) \phi =  \eR \phi      \qquad   |X| \le {a\ov 2}     \hfill  (\ref{eq8}b)^{\prime} \cr
 - {\hbar^2 \ov 2 m} {d^2 \ov X^2}  \phi =  \eR \phi      \qquad   |X|  > {a\ov 2}  }$$ 
  The eigen function of an even parity is 
\refstepcounter{equation}
$$  \displaylines{ \label{eq-phi}      
\hfill   \phi(X)= A \cos (K X), \qquad |X|\le {a\ov 2}   \hfill (\ref{eq-phi}a)
\cr   \hfill
  \phi'(X)= C e^{-K' |X|},   \qquad      |X| > {a\ov 2 }         \hfill (\ref{eq-phi}b)   }$$ 
where $\hbar K = \sqrt{2m (\eR +|\VRo|)}$ and $K'= {\sqrt{-2m\eR}\ov \hbar}$.  Applying the boundary condition at $X=|a/2|$, $\zeta \tan(\zeta) = K' (a/2)$, which, when substituted with the expressions for $ K$ and $K'$ above, writes as $\eta\equiv {\rm ctan} (\zeta)={\zeta \ov \sqrt{\beta^2-\zeta^2} }$, with $\zeta \equiv K (a/2)$ and $\beta \equiv {\sqrt{2m|\VRo| \ } \hbar} (a/2)$,
the energy solution is given to be 
\refstepcounter{equation} $$ \label{eq-Dq} 
\eR =- {\hbar^2 \zeta^2   \ov 2 m (a/2)^2 \eta^2}.   \eqno(\ref{eq-Dq}) $$ 
 Starting with a guess value for $\VRo$ and thus for $\beta$, put $a=a_r=2.64$ \AA \ for the same reason as before, solve graphically for $\eta$ and $\zeta$, and thus $\eR$; use the resulting $\eRo$ value in (\ref{eq-D0}) (below) and $\Do$  in turn  in (\ref{eq-Pp}) to acquire a new $\VRo$ value; and execute this iteratively until  $\VRo$ reaches constancy. We find, for  $\VRo$ being small here there is only one, hence the ground state solution: 
\begin{eqnarray} \label{eq-VRo}
\VRo=-10.65 \ (^o{\rm K}), \qquad  {\rm thus} \ \beta = 1.749,  \nonumber \\ 
 \eta_0= 0.68, \quad \zeta_0 = 0.31 \pi. \end{eqnarray} 
 With the $\eta_0$ and $\zeta_0$ values above in (\ref{eq-Dq}) we get the  ground state eigen energy of the $R$ motion:
$$ \eRo  =-7.13  \quad ^o{\rm K/atom} \eqno(\ref{eq-Dq})' $$
With the $ \eRo $ we further get:
\begin{eqnarray} \label{eq-D0}  \Do = \eRo - \VRo   = 3.45 \quad ^o{\rm K/atom}    \end{eqnarray}
The solution of odd parity is discarded as it does not fulfil the requirement of a symmetric (ground state) wave function for identical boson particles (Sec. \ref{Secz5.2}-2); see further Sec. \ref{Secz5.6}. 

\subsection{Momenton oscillation and excitation of collective phonons } \label{Secz5.3} 
\subsubsection{Set-up of Eqs. (\ref{eq9}a)-(\ref{eq9}b) and the solutions} \label{Secz5.3.1}
Equations (\ref{eq9}) are formally set up for every momenton masses, $M_j$, $j=1,2, \ldots$, with the inter-momenton potential $\V(\Xib+\xib)$, in the independent harmonic oscillator approximation.  
Since the momenton contribution to the total energy and excitation is negligible (see below), we will not solve its exact motion here and will only look at its long wave propagation, using an average mass $M$ and with an effective force constant $\alpha_2$. Equations (\ref{eq9}) thus reduce to a single equation similar to Eq. (\ref{eq7}a)$'$; its energy solution is thus:  
\begin{eqnarray} \label{eq2-7bm} \E_2(K_M) \approx \hbar c_2 K_M, \end{eqnarray}
where the wave number and velocity of a collective phonon are $K_M = 2\pi / \prod$, $ \La_M$ being the momenton wavelength,
and
\begin{eqnarray} \label{eqc2} c_2 = {\pd (\E_M/\hbar) \over \pd K_M } = \sqrt{{\alpha_2 \ov M}} \ L_M = \sqrt{{\alpha_{21} \over m}} \quad a, \end{eqnarray}
respectively. Apparently $c_2$ corresponds to the velocity of second sound discussed in Sec. \ref{Secz2.3}; $\alpha_2$, $ M = (N_{M})^3 m $, $L_M = N_M a$ and $N_M$ are the restoring force constant, mass, dimension and number of helium atoms in one dimension of a momenton oscillator, respectively. $ \alpha_{21} \simeq \alpha_2 /N_M^3$ represents then the effective force constant between two He atoms at the interface of two momentons.  
 $c_2(T)$ is dependent of $T$ via:
\begin{eqnarray} \label{eq-c2} {d c_2(T) \over d T} = {\partial c_2 \over \partial N_M}{\partial N_M \over \partial T}+ {\partial c_2 \over \partial \alpha_{21}} {\partial \alpha_{21} \over \partial T}. \end{eqnarray} 
Suppose at a higher $T$ a momenton will most likely find itself immediately surrounded by normal fluid regions, the reduction of $T$ is thus mainly to increase the number of momentons. Then $N_M$ and $\alpha_{21}$ and, by Eq. \ref{eq-c2} hence $c_2(T)$, are basically constant. Meanwhile, the interaction of a momenton with the surrounding normal fluid gives: $\alpha_{21} < \alpha_{1}$.  
Immediately this gives: $c_2 < c_1$.  As $T$ is reduced (to about 1 \oK) \ such that with a falling $T$,
each existing momenton increases in both $L_M$ and in its adjacency with other momentons for which $\alpha_{21}$ enhances, i.e. 
${\pd N_M \ov \pd T}<<0$ and ${\pd \alpha_{21}\ov \pd T}<<0$. Hence, by Eq.  (\ref{eq-c2}) ${d c_2 \ov dT} <<0$. In other words, $c_2$ increases rapidly with falling $T$, this being consistent with the experimental observation discussed Sec. \ref{Secz2.3}. Furthermore, as with Eq. (\ref{eq2-8}), the momenton excitation energies are $ {\LARGE \in}_{\Nc} = (\Nc+1/2) \E_M(K_M)$, where $\Nc=0, 1, \ldots$, and $ \E_M(K_M)= \hbar \Omega(K_M)$ is the energy quantum of a collective phonon.  Typically $\E_M(K_M) \sim 7.6 \times 10^{-9} $ \oK, \ as based on the upper-limit in the experimental second sound wave frequency, $\Omega \sim 10^3$ cycle/sec. Hence, $\E_M(K_M)$ is negligible compared to the typical energy scale of a single phonon, $\hbar \omega \sim 0.8$ \oK \ ($\sim 10^{11}$ 1/sec). Accordingly the momenton oscillation amplitude $|\Phi_{\xi}(\xi)|^2$ would be basically constant on the scale of $|\varphi(u)|^2$ and can be dropped from (\ref{eq-psi}). Finally, since $\La_M$ is large, a momenton wave will be absent in a small sample.

\subsubsection{Momenton wave, the atomic dynamics of second sound wave and temperature fluctuation} \label{Secz5.3.2}
Consider a given normal-mode frequency $\W$ of momenton oscillation that is equal to the normal-mode frequency $\w$ of single-atom  oscillation; the latter consists of all wavelengths (much shorter than the momenton dimension) within the momenton. Two extreme cases can present. With some momentons, the $\W$ periodic process is in phase with the $\w$ oscillation. Evidently, the $\W$ wave of amplitude $A_{\W}$, which will superpose with the $\w$ wave of amplitude $A_{\w}$, will then increase the $A_{\w}$ to $A'_{\w}=A_{\w}+A_{\W}$. And conversely with some other momentons. Therefore, in each of the former momenton regions the phonon population $n$ and thus the temperature $T$ are increased by small amounts $n'-n \propto 2A_{\W}A_{\w} $ and $T'-T$, as $(A_{\w})^2 \propto n$,
noting $A_{\W}<<A_{\w}$; and conversely in the latter regions. So that a temperature wave forms, which is one defining characteristic of the so-called second sound wave.\cite{Ash}

\subsection{ Superfluid flow, translation of momentons or of single atoms as of Eqs. (\ref{eq10})-(\ref{eq10}b)} \label{Sec5.5}

Equation (\ref{eq10}a) describes a steady flow motion of the superfluid in the absence of convection and is expressed for the translation of the equilibrium position $\Xib$ of a momenton.  Under normal friction conditions, the flow is driven by an external pressure difference $P_1-P_2$ acting on a cross sectional area $S$ of the flow, hence $\FS=(P_1-P_2)S$.  One portion of the $\FS$ is used to overcome the viscous resistance from the channel wall $f$ acting on the flow through a flow-wall contact area $\Asw$;  and the remaining to accelerate the flow inertia, the flow mass, which is
\begin{eqnarray} & & M_{s} = \sum_i M_i=N_{s}m. \end{eqnarray}
$N_{s} $ is the total number of helium atoms in the flow, and $m$ -- as before -- the atomic mass of helium. Because $ <\ub> = <\xib> =0 $ and also $<\Rb> \approx $ constant ($\bf C$) ($< >$ denotes the time average of a variable over an infinitesimal flow translation), 
therefore
\begin{eqnarray} \label{eq11} <\rb> (t) \ = \Xib (t) \ +{\bf C}. \end{eqnarray}
(\ref{eq11}) implies that the translation of a single atom, or a momenton, or the flow, describes the same motion. 
If  $f \approx 0$, which will occur when $v_s$ is below a critical value (which criterion is given in paper II), Eq. (\ref{eq10}a) then describes the non-dissipative superfluidity flow motion. 
$\FS$ is then entirely consumed to accelerate the flow velocity, for example at time $t=\tau$, to  $v_s(\tau)= (1/M_s) \int_0^\tau \left[ \int_0^\tau (P_1-P_2)S (t) dt  \right] dt$.  If from $t \ge \tau$, the driving pressure is removed: $(P_1-P_2)|_{t \ge \tau}=0$. Then Eq. (\ref{eq10}) and its solution become simply 
$$ M_s {\pd ^2 <\Xib>\ \ov \pd t^2} = m {\pd ^2 <\rb>\ \ov \pd t^2}= 0,  \quad \quad {\rm and } \eqno(\ref{eq10}a)' $$
\begin{eqnarray} \label{eq2-13b} v_s = {\partial <\rb>\ \over \partial t}= {\partial <\Xib>\ \over \partial t} = v_s(\tau) \equiv {\rm constant}. \end{eqnarray}
The classical equation (\ref{eq10}a) validly describes the motion of the center of mass of an atomic wave packet. 

The translations of the quantum helium II atoms are more fully, or strictly, described by Eq. (\ref{eq10}b). As $\HS=p^2/2m + \VS$ and $\VS=$ 0  in the absence of non-uniform external field, the solution of (\ref{eq10}b) is an "effective" planewave:
\begin{eqnarray} \label{eq2-13c} {\mit \Phi} (<\rb>) = C e^{\imath({\bf k_s} <\rb> - \omega t)}, \end{eqnarray} 
with "effective" referring to the fact that each single atom, if driven externally, is accelerated according to the relation (\ref{eq10}), 
to a velocity $v_s$ with an effective mass equal to the flow mass $M_s$.   The wave number $k_s$ is connected with $v_s$ by the de Broglie relation:
\begin{eqnarray} \label{eq2-13d} k_s = m v_s / \hbar; \end{eqnarray}
$k_s$ is thus similarly effective. The translation kinetic energy of an atom is accordingly 
 $e_s =mv_s^2/2 = \hbar^2 k_s^2/2m$. 
The flow energy is thus: $E_s= {1\ov 2}M_sv_s^2 =N_s e_s =N_s {\hbar^2k_s^2\ov 2m_s}$.  
It is noteworthy that, Eq. (\ref{eq2-13d}) holds only for the steady flow motion here, where 
periodical processes of each atomic waves are essentially linear, independent and hence obey the 
 superposition principle of quantum mechanics. However, when some part (e.g. the interfacial layer) of the flow is being resisted as opposed the remaining part, the former will interact with the latter  given that the superfluid atoms are strongly correlated. Then, each atom in respect of its flow motion is no longer a "free" particle.  

  By knowing $v_s$ and the density $\rho_s = |\psi|^2$, with $\psi(r)$ as in Eq. (\ref{eq-psi}) 
the superfluid current density is given accordingly
\begin{eqnarray}  \label{eq3-5}
\jb_s = {\hbar \over \imath 2m} [\psi^* \nabla \psi -(\nabla \psi^*) \psi ] = \rho_s v_s. \end{eqnarray} 
For the two-fluid coexistent of He II under thermal equilibrium, by Sec. \ref{Secz2.3} each of the fluids are maintained in their own regions with no intermingling on an atomic scale, therefore each fluid maintains its specific density.  For a steady, non-compressible flow, the continuity conditions follow accordingly
\begin{eqnarray} \label{eq3-6} 
 {\bf \nabla } \cdot  \jb_s + { \pd  \rho_s \over \partial t } =0  \qquad {\rm and} \qquad 
 {\bf \nabla} \cdot \jb_n + { \pd \rho_n \over \partial t }=0. \end{eqnarray}
For a steady superfluid flow in a straight channel, the above becomes $ {\bf \nabla } \cdot \jb_s = { \pd  \rho_s \over \partial t }=0$ and $ {\bf \nabla } \cdot \jb_n = { \pd \rho_n \over \partial t}=0$.  For a steady flow of a two-fluid mixture maintained by a pressure difference there is $v_s=v_n$, and the island-like composite structure 
(FIG. \ref{fig-mom}a)
should then maintain. If now the driving pressure is removed, so that, starting from the layers by the wall, $v_n=0$ and only the (pure) superfluid will flow.  A natural course via which this actually takes place is that the superfluid will break through the normal fluid and re-shape itself into narrow strips.

\subsection{The total and full excitation spectrum, the total wavefunction and total zero point energy} \label{Secz5.6}

\paragraph{The total and full excitation spectrum.} The total excitation energy for a superfluid at rest is 
\begin{eqnarray} \label{eq-ev} &  \ev(K) & = \E(K)+ \Db(K) + O(\E_M) \nonumber \\
                           & & \simeq \E(K)+ \Db(K).\end{eqnarray} 
 The full spectrum of $\ev(K)$ in (0, 1.93) \AAi \ is combinatorially given by Eqs. \ref{eq2-7b}a, b,c,d (solid curve 1, the dashed curve, and solid curve 2  in FIG. \ref{fig-wVK} and also in FIG. \ref{fig-ex-spec}). The single phonon and the superfluid bond excitations are, as the evaluation of specific heat in Sec. \ref{Secz9} will show, characteristic in the $T$ regions (0, 0.6) and (0.6, 2.17) \oK \ respectively.

\paragraph{The total zero point energy} is: 
\begin{eqnarray} \label{eq-ez} 
& \exists_0 &  = \e_0 + \Do + O(\in_0) \nonumber \\
& &   \simeq 7.5 + 3.45 = 10.95 \ ^o{\rm K/atom} 
 \end{eqnarray}
where $\e_0$ and $\Do$  are given by (\ref{eq-ena}) and (\ref{eq-D0}). 

\paragraph{The total ground state potential energy} is:   
\begin{eqnarray} & \Ph0 & = V_0 -  \exists_0 = \VRo - \e_0 - O(\in_0) \nonumber \\
 & & = -7.5 -10.5 = -18.5  \ ^o{\rm K/atom }  \end{eqnarray}
where $\VRo$  is given by Eq. (\ref{eq-VRo}).

\paragraph{The total wavefunction.}  
Of the $\psi(\rb)$, Eq. (\ref{eq-psi}), for an atom at $\rb(x,y,z)$,  given the rapid $\ub$ oscillation, say in the $x$ direction, $\varphi (u(X))$ as of Eq. (\ref{eq-vph}) will  on the time scale of the $R$ motion show as an time average to be an even function relative to $X$. 
$ \Psi_{\xi}(\xib)$ is even both for a similar reason and because $ |\Psi_{\xi}|^2$ is essentially flat (and shallow)  within the region of wavefunction overlap. $\phi(X)$ of  Eqs. (\ref{eq-phi}a)- (\ref{eq-phi}b) is even. Last, $\Psi_{\Xi}(\Xib)$ is identical for all atoms in the (flow) bulk. 
Hence, overall, on the time scale of $R$ motion, by which the atoms may exchange positions on activation,  $\psi(x)$ is an even function with respect to the equilibrium position $X$; and $x$ can be an arbitrary direction relative to the isotropic fluid.  Therefore, $\psi(x)$ is spherically symmetric about $R$. The product of the single particle wave functions therefore yields a desirable, symmetric total wave function  of $N$ identical (in the sense of Sec. \ref{Secz5.2}-2) particles of the fluid bulk: 
\begin{eqnarray}\label{eq-Ptot} {\it \Psi} (r_1, r_2, \ldots, r_N) = \psi(r_1) \psi(r_2) \ldots \psi(r_N). \end{eqnarray}

 \section{The stability condition of the superfluid } \label{Sec6} \label{PSec4} \label{Secz6}

From the superfluid bonding character we can infer: 
1). The substantial superfluid bond, being larger than the van der Waals "bond" of He I (quantitative evaluation given in Sec. \ref{Secz9}), implies a substantial {\it internal viscosity}, larger than that of He I. Thus the superfluid is not an assembly of {\it non-interacting atoms} and hence cannot  be of {\it a zero internal viscosity} -- as portrayed in London's BEC and in the unified LLBF theory as well.  
Meanwhile, the "weakly" (or sometimes referred to as "substantially") interacting superfluid atoms" and the "zero viscosity" which are simultaneously held in the LLBF theory, are two inevitable self-contradictory aspects.  2). Now that the viscosity of liquid He I can sustain a non-turbulent macroscopic rotation (below certain velocity) of He I which is a commonplace experimental observation, the larger internal viscosity of the superfluid He II can therefore certainly ensure a (more robust) {\it non-turbulent} macroscopic rotation of He II.  The superfluid is thus not {\it irrotational}.  3). It therefore also follows that for the superfluid a {\it zero total circulation}, $\kappa_{tot}=0$, is not a necessary condition for non-turbulence; instead $\kappa_{tot}$ is finite and quantized, as has been experimentally demonstrated and to which a consistent theoretical grounding is provided in  Sec. \ref{Sec7}.  4). Accordingly, no {\it local vorticity} needs be imposed on the superfluid, as the Stoke's theorem, \cite{lamb} applying for a viscous-less fluid, does not apply to the superfluid.  5). It also follows from points 1)-3) that the grounds for proposing the "absence" of internal viscosity as an explanation for the non-dissipative superfluidity motion are lacking. (This is in addition to the fact that an internal viscosity does not necessarily imply which "flow-wall" viscosity, which we are however not to expand on here.)

\section{The quantization of circulation} \label{Sec7} \label{PSec5} \label{Secz7}

Consider a superfluid bulk in steady rotational flow motion with a tangential velocity $v_s$.  
The trajectory of a flow atom $<r>$, Eq. (\ref{eq2-13c}), can be now conveniently projected onto the circular coordinate $\ell$ along the flow streamline, $<>$ being here for simplicity omitted. 
The atomic wave, assuming being localized relative to the rotational frame and hence not to swapped off the stream line,  in the circular loop, of a circumference $\Lov$, must satisfy according to quantum mechanics the continuity condition: 
\refstepcounter{equation}
$$ \label{eq-nlmb} 
k_s \overline{L} = n 2\pi.   \eqno(\ref{eq-nlmb}a)
$$ 
This then yields a circulation quantization as to be expressed in Eq. (\ref{eq-circ3}) below. 
The case here is seen to be equivalent to an electron wave orbiting about the nucleus, except for their different centripetal forces.

Alternative to the above, to examine its microscopic detail we below look at the process from the standpoint of wave interference. We first determine an important parameter, the length of the rotational wave train of a helium atom, and this in theory is simply 
\begin{eqnarray} \label{eq-Lm} \Lw=\Nlmb \cdot \lambda_s  \end{eqnarray}
 where $\lambda_s$ is the wavelength and $\Nlmb$ is the number of $\lambda_s$'s of the rotational wave train. The de Broglie relation (\ref{eq2-13d}) informs about $\lambda_s = 2\pi/k_s$, and not the $\Nlmb$. As to the physical acquisition of $\Lw$, let us think that the action of passing a $\hbar k_s$ to a helium atom consists in passing it a field -- in the form of a (purely) rotational wave train -- whose momentum equals $\hbar k_s$. [We are tempted to think of this field being an electromagnetic wave, although this link would require a justification which goes beyond the context  of this paper.]
Since this wave is given as the solution of the Schr\"odinger equation (\ref{eq10}b), we shall merely take this consequence. We can then think of the total atomic wave train now rotating along $\ell$ as  the stationary thermal de Broglie wave train of a helium atom as being superposed in the rotational frame onto the purely rotational wave train of length $\lambda_s \cdot \Nlmb$. The former, having a wavelength of typically nano meters, will not contribute to the interference of a macroscopic length period. Consider now only the latter; its $\Nlmb$ is evidently determined by the interaction time, $\tauw$, 
taken for creating the purely rotational wave train. That is, $\tauw= \Nlmb T_s$, $T_s=2\pi/\w_s$ being the time period of the wave (or the period of its generating source), and $\w_s =2\pi v_s/\lambda_s$. 
Clearly, for wave interference to occur at a low $n$ number in Eq. (\ref{eq-nlmb}), first of all $\lambda_s$ needs be so large as to be comparable with $\Lov$. 
Given $m$, of the helium atom here, is large compared to that of the usual quantum particles such as the electron, 
an (arbitrarily) large $\lambda_s$ ($=\hbar /mv_s$) can be achieved by compensatorily making the flow velocity $v_s$ (arbitrarily) small, thus the $T_s$ and $\Lov$ being accordingly large.
In experimentation, for instance, a large $T_s$ value ranging $30 \sim 600 $ seconds, with a $\Lov \sim $ 0.0025 cm were achieved.\cite{vinen} Furthermore, 
to achieve $n=\Nlmb >2$ or $>>2$, the $\tauw$ should $>2 T$ or $>>2T$. 

If the above conditions are met. We can then achieve $\Lw=\Nlmb \lambda_s  \ge 2 \Lov$, with $ \Nlmb \ge 2$.  In this way, each rotational atomic wave train will wind about each loop at least twice, interference will now occur.  For the loop satisfying  
$$ \Lov= 2n {\lambda_s \over 2}, \eqno{(\ref{eq-nlmb}b)}$$
where the $\Lov$ will sometimes also write as $\Lov_{2n}$ and $n$ is as before an integer, then, the rejoining waves will be in phase:  
\begin{eqnarray}\label{eq-cir-3.a}
& & {\mit \Phi} (\ell_1' ) = e^{\imath (k_s \ell_1 + \oint k_s d\ell)} = e^{\imath (k_s \ell_1 + 2\pi n)}  \\
& & = e^{\imath k_s \ell_1} ( \cos 2 \pi n + \imath \sin 2\pi n) = {\mit \Phi} (\ell_1). \nonumber  \end{eqnarray}
This gives an amplified total amplitude: $\Psi (\ell_1' )+\Psi (\ell_1)=2\Psi (\ell_1)$. 
Re-writing Eq. (\ref{eq-cir-3.a}) in terms of circulation, $\krot$, we get:
\begin{eqnarray} \label{eq-circ3} \krot = \oint  v_s  d \ell = {\hbar \over m} \oint k_s d\ell = {h \over m} k_s {\overline L} = { nh \over  m}, \nonumber \\ \quad \quad n=0,1,2, \ldots. \end{eqnarray}
that is, the circulation is quantized.
Or, if $ \Lov = (2n+ 1) {1 \over 2}\lambda_s$, the amplitudes will cancel out: 
${\mit \Phi}(\ell_1)+{\mit \Phi}(\ell_1')=0$.  
The amplitudes will partially cancel if $\overline{L}$ is a fraction of $\lambda_s$ and $n$ is small ($\sim 2$),  but will cancelled out quite cleanly if $n >>2$ in which case the partially out-of-phase waves are superposed a large $n$ number of times. The above is a standing wave to the observer who rotates with the flow.  Since all of the rational atom waves in the flow have identical $v_s$, $m$, and hence identical $\lambda_s$, and are also in phase (for they are set to rotation at the same time and they are relatively localized), thus they will all present amplified amplitudes at the same circumferences $\Lov$ defined by (\ref{eq-nlmb}b), and zero elsewhere. In the former case, the linearly superposed atomic waves yield a total enhanced fluid density, presenting macroscopically discrete streamlines. 

Finally, by Sec. \ref{Secz6}, the macroscopic rotation following the vessel above is the only rotational motion, therefore the total circulation is $\kappa_{tot} = \kappa_{rot}= nh / m$, where $n=0,1,2, \ldots$. The signal of a measurement, if being in proportion to the total amplitude of the atomic waves, will then inform a circulation quantization.

The physical process described above, which can be termed the {\bf circular atomic wave self-interference}, provides a mechanism for the circulation quantization that has been experimentally demonstrated.\cite{vinen,Rayfield:1964,Whitmore-Zimmermann:1968,Donnelly} 

\section{The dynamic structure factor about $q_b$ as due to the superfluid bond excitation} \label{Sec8} \label{Secz8}
Corresponding to the total excitation energy $\ev(K)$ (\ref{eq-ev}), the total dynamic structure factor from a (pure) superfluid is  
\refstepcounter{equation} 
$$\displaylines{ \label{eq-Stot}
\hfill S(q,\omega) = S(q, [\E(K) + \Delta(K)]/\hbar)   \hfill  (\ref{eq-Stot})  \cr
  = \left\{ \begin{array}{cc}  
          S_{\ph}(q,\E(K)),   \quad    q<q_b-{\sigma \ov 2}    & \qquad\qquad\mbox{ (a)}   \cr 
         S_{\ph, b}(q, [\E(K-K_b)+ \Db]/\hbar),   \quad    |q-q_b|<{\sigma \ov 2} & \qquad \qquad \mbox{  (b)} \cr
         S_b(q, \Db),   \quad    q=q_b          & \qquad\qquad\mbox{(c) }
\end{array} \right. 
}$$
here $\sigma $ represents a narrow vicinity of $q_b$. 
 For the phonon excitation, $S_{\ph}(q,\omega)$, by Eq. (\ref{eq2-8}) the $ \e_n$'s of the scattering system are discrete.
Neglecting phonon-phonon collisions, which is to a good approximation as based on neutron scattering measurements (Sec \ref{Secz2.1}) and thermal conductance measurements, \cite{Fairbank-Wilks} and as has also been theoretically evaluated by Cohen and Feynman,\cite{Feynman:1957}  it  follows then effectively
\begin{eqnarray} S_{\ph}(q,\omega) \propto \delta\left(\omega(K) -{1\over \hbar}\left(\epsilon_n(K)-\epsilon_{n+1}(K)\right)\right), \end{eqnarray} 
where $K=q (\equiv |\kne'-\kne| )$, $\kne$ and $\kne'$ being the wavevectors of a neutron before and after scattering. 

At  $q_b=2\pi/a=b$, $b$ being reciprocal pseudo lattice vector, by Sec. \ref{Secz2.2} $S(q,\omega)$  is due to the excitation of a superfluid bond and is thus expressed by (\ref{eq-Stot}c). 
In $|q-q_b|<\sigma/2$ where when $S_b(q, \omega)>>S_{\ph, b}(\w,q)$, then (\ref{eq-Stot}b) writes $S_{\ph, b} \approx S_b(q, \w)$; a switching-over region where $S_b(q, \omega)$ and $S_{\ph, b}$ may have similar weights  but would be very narrow, and is not to be dealt with here.   
As Sec. \ref{Sec2.2} reviews, $S_b(q,\omega)$ as measured by neutrons presents a peculiarly high and sharp peak centered at $\w= \Delta_s/\hbar =8.6 $ \oK. \ 
Below we formally express $S_b(q,\omega)$ through a few combinatory modifications from a crystalline solid scattering for which we have the precise description, to elucidate the route via which the superfluid has acquired this particular scattering feature, apparently differing from that of either a solid or a liquid.

(a). Let us start with an imaginary superfluid, in which the atoms are arranged in a periodic lattice of a simple cubic structure, and the atomic bonding and activation energies, denoted by $\Delta_b^{\cry}$ and $\Delta_s^{\cry} = \Delta_b^{\cry}+ \delta^{\cry}$, are as large as in a solid. For this system, (only considering the longitudinal modes) the excitation energy $\E(q)$ is regularly given by Eq. (\ref{eq2-7b}) to be a sinusoidal dispersion, the dotted line in FIG. \ref{fig-ex-spec}a, 
which is zero at $q_b = {b}$. $\E(q_b)=\hbar \w(q_q)=0$ together with  $q_b a=2\pi $ implies that at $q_b$, $S_{\ph}(q_b,\omega) \equiv 0$ and an elastic Bragg scattering occurs. The scattering intensity at a given $q$ is proportional to the static structure factor, which (strictly) at zero \oK \ is
\refstepcounter{equation}
$$\displaylines{ \label{eq-sc} 
\hfill   S^{\cry}(q)  = {1 \over N} \sum_{i,j}^N e^{\imath \qb \cdot (\rb_i - \rb_j)}  \hfill \cr
\hfill \qquad  = \int \left[ S_{\ph}(q, \omega)  + S^{\cry}_b(q, \omega)\right] \delta(\omega) d(\hbar \omega)  \hfill   \cr
\hfill \qquad  = \hbar S_b^{\cry}(q,\omega)|_{\omega=0}.  \hfill  (\ref{eq-sc}a)
} $$
The first expression of (\ref{eq-sc}a) is just the Bragg scattering condition; it gives $S^{\cry}(q)=1$ at $\qb \cdot (\rb_i - \rb_j) = q_b a = n 2\pi$ $(n=0, 1,$ \ldots) and zero elsewhere.  
The last two expressions of (\ref{eq-sc}) generalize $S^{\cry}(q) $
to include an energy transfer $\w$ but with $\w=0$ for the {\it elastic} scattering of the present system. The {\it elastic} feature is readily understood to result principally from the inequality:
\refstepcounter{equation}
$$ \label{eq-sc11} 
{{\hbar^2 k^2_{\neu} \over  m_{\neu}} \atop (5\sim 100 \ {\rm meV})} {<< \atop } 
{\Delta_s^{\cry}, \atop (1 \sim 10 \ {\rm eV \ for \ solids})}  \eqno(\ref{eq-sc11}a)
$$
where $ k_{\neu}$ ($=m_{\neu}v_{\neu}/ \hbar$), $v_{\neu}$ and $m_{\neu}$ are the velocity and mass of incident neutrons.
(b). We now reduce $\Delta_b^{\cry}$ ($= \Delta_s^{\cry} - \dv{}^{\cry}$) to equal the superfluid bond $\Delta_b$ ($ = \Delta_s - \dv$) $\simeq 8.6 $ $^o$K (0.74  meV), but retain the periodic configuration. 
Then 
$${\hbar^2 k^2_{\neu} \over 2m_{\neu}} >> \Delta_s.  \eqno(\ref{eq-sc11}b)$$ 
Thus $\Delta_b$ can readily be excited by neutrons. 
Then, Eq. (\ref{eq-sc}a) is no longer, in principle, an exact description for the present {\it inelastic} scattering.  However, supplementing the condition Eq. (\ref{eq-sc11}b) to the conservation relations for the inelastic scattering at $q=q_b$:
\input epsf  
\begin{figure} [here]  \begin{center} \leavevmode \hbox{%
\epsfxsize= 7.5 cm     \epsfbox{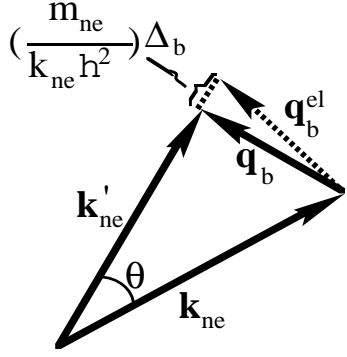}} \end{center} 
\caption{   Schematic diagram of the relationship between neutron wavevectors $k_{\ne}$ and $k'_{\ne}$ before and after an inelastic scattering, and the momentum loss $q_b$ that is associated with the energy loss $\Delta_s$. The Bragg condition is approximately met given that $\Delta_s <<e_{\ne}$ (the incident neutron kinetic energy) and $k_{\ne}'-k_{\ne} \approx \Delta_s / \hbar e_{\ne} << k_{\ne}$, therefore $k_{\ne} \approx k'_{\ne}$. The closed vector-loop ending with $q_b{}^{el}$ represents the exact Bragg condition. \label{fig-scat}  } 
\end{figure}
\noindent
\begin{eqnarray} \label{eq-brag1} | \kb_{\neu}' - \kb_{\neu} |  = q_b = {2\pi \over a },  \\ \label{eq-brag2} 
{\hbar^2 \over 2 m_{\neu}} (k'^2_{\neu} - k^2_{\neu})=\Delta_b, \end{eqnarray}
 we obtain that (\ref{eq-brag2}) can be adequately approximated by 
$$ k'_{\neu} \simeq k_{\neu}; \eqno(\ref{eq-brag2})'$$
  see FIG. \ref{fig-scat}.
Taken together, (\ref{eq-brag1}) and (\ref{eq-brag2})$'$ effectively yield the Bragg condition
\begin{eqnarray} q_b \simeq 2 k_{\neu} \sin \theta = {2 \pi \over a}, \end{eqnarray}
which is equivalent to Eq. (\ref{eq-sc}). In other words, (\ref{eq-sc}a) effectively holds correct for the periodically ordered superfluid, but with $\omega=0$ replaced by $\w = \w_s $ $=\Delta_s/\hbar$. 
Modifying (\ref{eq-sc}a) accordingly, we have
\begin{eqnarray}  
&& S^{\lat}(q)|_{\w= \Ds/\hbar}  \dot = {1\over N} \sum_{i,j} e^{i \qb(\rb_i - \rb_j)} \delta(\omega-\Delta_s/\hbar) \nonumber \\
&& \dot =\int [S_b(q,\omega) + S_{\ph}(q,\omega)] \delta(\omega-\Delta_s/\hbar) d(\hbar \omega) \nonumber \\
&& =  \hbar  S_b^{\lat}(q, \omega)|_{\omega=\Delta_s/\hbar}. \qquad (q \in q_b \pm \sigma)
\quad\quad\quad\quad\quad (\ref{eq-sc}b) \nonumber \end{eqnarray}
Re-ordering (\ref{eq-sc}b), we get
\begin{eqnarray}  \label{eq-Sbc}
 &&S_b^{\lat}(q, \omega)|_{\omega=\Delta_s/\hbar} = {1\ov \hbar}S^{\lat}(q)|_{\w= \Ds/\hbar} \nonumber \\
 &&=  {1\over N \hbar } \sum_{i,j} e^{i \qb(\rb_i - \rb_j)} \delta(\omega-\Delta_s/\hbar)  \qquad (q\in q_b) 
\end{eqnarray}  
This says that  $S_b^{\lat}(q, \omega)$ is a modified $S^{\lat}(q)$, and is maximal when $q=q_b$ and $\omega = \Delta_s/\hbar$.  To re-emphasize, the acquisition of Eq. (\ref{eq-sc}b) is a result of $\Delta_b$ being so negligibly small -- of the typical energy scale of a phonon -- compared to the incident neutron energy that it causes minimal departure from the Bragg condition; $\Delta_b$ is nevertheless sufficiently large to be detectable using the thermal neutrons. This incident is analogous to the thermally broadened Bragg scattering -- damped by a Debye-Waller factor -- in solids.

(c). We finally return from (b) to the real superfluid by disordering the atoms. 
A well-defined quantity is now, as with any liquid, 
the mean interatomic spacing $a$ at which the radial distribution function $g(r)$ peaks. As is commonly done with regular liquids \cite{Egelstaff,Bacon,Squires} or quasi-crystals,\cite{Janot} we can replace the site summation of Eq. (\ref{eq-Sbc}) -- invovling a $\delta$-function -- with an integral involving $g(r)$ and meanwhile replace the $\delta(\w-\Delta_s/\hbar)$ with $h(\omega-\Delta_s/\hbar)$; further combining with  (\ref{eq-Stot}b)-(\ref{eq-Stot}c), we get:
\begin{eqnarray} \label{eq-Sb-c}       
&&S(q, \omega)|_{\omega \sim \Delta_s/\hbar}= S_b(q, \omega)|_{\omega \sim \Delta_s/\hbar}  \nonumber\\
&& ={1\ov \hbar } \lf. S^{\lat}(q)\rt|_{\delta(r) \rightarrow g(r) \atop {\delta(\w-\Delta_s/\hbar) \rightarrow h(\omega-{\Delta_s\ov \hbar})}} \nonumber \\
&& ={1\ov \hbar } \left[1 + \int (g(r)-\rho) e^{i\qb \rb} d\rb  \right] h(\omega-\Delta_s/\hbar),  \nonumber \\
&&\qquad\qquad\qquad\qquad\qquad\qquad\qquad q \in q_b \pm \sigma
\end{eqnarray} 
here $h(\omega-\Delta_s/\hbar) = (1/\sqrt{\pi} \sigma)\int e ^{-[\hbar (\omega - \Delta_s/\hbar)/\sigma]^2} d\omega $ sharply peaks around $\Delta_s/\hbar$ and has a finite width $\sigma$. $\sigma$ ought to be basically determined by the fluctuation of the atomic bond length due to the short range ordering. 
(\ref{eq-Sb-c}) is the scattering function due to the superfluid bond excitation. Fixing $\omega \simeq \omega_s$, we expect to see $S(q,\omega)$ to peak on the $q$ axis; although, there is no experimental data available to us for comparison. 
Two further remarks can be made.  First, insofar as the "Bragg" scattering is concerned, it is only meaningful to compare the $S_b(q,\omega)|_{\w=\w_s}$ of Eq. (\ref{eq-Sb-c}) with  the zero temperature $S^{\cry}(q)$ of  Eq. (\ref{eq-sc}a)  and not with the finite temperature $S^{T}(q) (= \int^\infty_{- \infty}  S^{T}(q, \omega) d \omega)$ of the superfluid;  the $S^{\cry}(q)$ of a solid at zero \oK \  only involves the $\omega=0$ scattering, whilst the $S^T(q)$ of the superfluid contains scatterings of all $\omega$.  Second, the correspondence between  (\ref{eq-sc}) and (\ref{eq-Sb-c})  not only exists in a static sense, but also in a dynamic sense, as atoms in the superfluid, viewed in $\xib$, are localized as in a solid except that they present no long range ordering. As a basic difference from that of a normal liquid, on the other hand, the finite peak width in $g(r)$ of the superfluid is primarily due to the disordering of the atomic equilibrium positions but not atomic diffusion.

Through steps (a)-(c), the neutron scattering from the superfluid about $q_b$ emerges to be an {\bf inelastic damped-Bragg scattering} due to the excitation of the superfluid bond, formally described by Eq. (\ref{eq-Sb-c}).

\section{Thermodynamics of He II } \label{Secz9}
\subsection{The partition function of the respective fluids  } \label{Secz9.A}
We consider a bulk of superfluid He II in thermal equilibrium at $T$, having a volume $V$ and number of atoms $N$.  Employing the canonical ensemble of the independent variables $(T,V,N)$, the total partition function of the fluid formally is
\begin{eqnarray}  \label{eq3-11} Z_{\alpha} =   e^{{- \Ph0 \over k_B T }} {\tilde Z}_{\alpha}, \end{eqnarray} 
where $\a=s$ for the (pure) superfluid and $n$ for the normal fluid of He II, $\Ph0$ is the ground state potential energy and $e^{- \Ph0/k_BT}$ the corresponding partition function, and ${\tilde Z}_{\alpha}$ is the total kinetic energy partition function.  If the $ Z_{\a}$ is known, the thermodynamic functions, the internal energy $U_{\a}$, free energy $A_{\alpha}$, entropy $S_{\alpha}$ and specific heat $C_{V \a}$ can then be obtained from the standard relations: 
\begin{eqnarray} \label{eq3-12}
U_{\a} = k_B T^2 {\pd \ln Z_{\a} \over \pd T}
   = \Ph0 + k_BT^2 {\pd \ln {\tilde Z}_{\a} \over \pd T, } 
\end{eqnarray} 
\begin{eqnarray} \label{eq3-13}
&  & A_{\alpha}   = - k_B T \ln  Z_{\alpha} = \Ph0  - k_BT \ln  {\tilde Z}_{\alpha},  \\
\label{eq3-14}
&  & S_{\alpha}  = - { \pd A_{\alpha} \over \pd T } =  k_B  {\pd \left( T \ln {\tilde Z}_{\alpha}  \right) \over \pd T}= {1 \ov T} (U_s - A_s),\\ 
\label{eq3-15} \label{eq-CV0}
& & C_{V \alpha } = \left({\pd U_{\alpha} \over \pd T}\right)_{V,N}.  
\end{eqnarray} 
The microscopic theory laid out in the preceding sections enables us 
to treat the two fluids as independent, macroscopic systems, and the practical scheme to obtain in the following sections \ref{Secz9.B}-\ref{Secz9.D} the respective partition functions and the thermodynamic functions.

\subsection{The pure superfluid }  \label{Secz9.1}   \label{Secz9.B}
Consider a pure superfluid of $N$ helium atoms, which are by Sec. \ref{Secz2} relatively localized, in the normal coordinates can be represented by $N$ independent, identical and distinguishable (see the discussion in Sec. \ref{Secz5.2}c), effectively harmonic oscillators. With the single oscillator kinetic energy $\e_n (K) $ of Eq. (\ref{eq2-8}), we get the kinetic partition functions for single oscillator:
\begin{eqnarray} \label{eq3-16}
& z_{\ph}(K) & = \sum_{n=0}^{\infty} e^{{\e_n(\w) \ov k_B T} }= \sum_{n=0}^{\infty} e^{- (n + {1\ov 2}) {\hbar \omega (K) \ov k_B T} } \nonumber \\
& & = { e^{ - \hbar \omega(K) / 2 k_B T } \over 1 - e^{- \hbar  \omega(K) /k_B T }}, \end{eqnarray}
and for $N$ oscillators: $\Pi_{K} z_{\ph}(K)$. Similarly the partition functions for a single and for $N_2=\g N$ momentons are as Eq. (\ref{eq3-16}) and $ \Pi_{K_M} z_{cph}(K_M)$, with $\W$ in place of $\w$. With the above we get
\begin{eqnarray} \label{eq3-18}
 {\tilde Z_{s}}  =  \Pi_{K} z_{\ph}(K)  \Pi_{K_M} z_{{\rm cph}}(K_M).  
\end{eqnarray}
Substituting Eq. (\ref{eq3-16}) and its momenton counterpart in Eq. (\ref{eq3-18}) and in turn,  (\ref{eq3-18}) in Eqs. (\ref{eq3-12})-(\ref{eq3-13}), employing the identity $\ln [\Pi_K z_{\ph}(K)  \Pi_{K_M} z_{{\rm cph}}(K_M)] = \sum_{K}  \ln z_{\ph}(K) + \sum_{K_M}  \ln z_{{\rm cph}}(K_M)$,  we thus obtain for the (pure) superfluid: 
\begin{eqnarray} 
& &U_s   
     =\Ph0 + k_BT^2 \lf[\sum_K {1\ov z_{ph}} {\pd z_{ph} \ov \pd T}+ \sum_{K_M} {1\ov z_{cph}} {\pd z_{cph} \ov \pd T}\rt]      \nonumber \\
  &    &      =  \Us0 +
  \int^{\w_D}_{0}  \hbar \w \D(\w) <n(\w)>  d\w + \OUM(\W)  \nonumber \\
& &  \simeq  \Us0 + {3Nk_BT^4 \ov \theta_D^3} \int^{y_D}_0 {y^3 \ov e^y-1} dy   \quad  (\OUM \approx 0)    \nonumber  \\
&   &  \simeq  \Us0 +   {Nk_B \pi^4  T^4   \ov 5 \theta_D^3},    
\qquad\qquad\qquad\qquad\qquad\qquad (\ref{eq3-12}a) \nonumber
\end{eqnarray}
\begin{eqnarray} \nonumber
& & A_s = \Us0 + k_B T  \int  \ln \left(1-e^{-\hbar \omega /k_BT}\right) \D(\omega) d \omega  
+O_{AM}(\W)    \nonumber  \\ 
&& = \Us0 + {Nk_BT^4 \ov \theta_D^3} \int^{y_D}_0 y^2 \ln (1-e^{-y}) dy \nonumber  \\ 
&& = \Us0 + {Nk_BT^4 \ov 3 \theta_D^3} \lf[ \lf. y^3  \ln(1-e^{-y})\rt|^{y_D}_0 - \int^{y_D}_0  {y^3\ov e^y-1} dy      \rt] \nonumber  \\ 
& & \simeq \Us0 - Nk_B {\pi^4 \ov 15} {T^4 \ov \theta_D^3}   
 \nonumber \\
   &&      \nonumber \qquad\qquad\qquad\qquad\qquad\qquad\qquad\qquad\qquad\qquad (\ref{eq3-13}a) 
\end{eqnarray} 
and simply with the $U_s$ and $A_s$ above in the last expression of (\ref{eq3-14}), we get
$$  S_s \simeq   N k_B   {2 \pi^4 \ov 15} \left( {T\ov \theta_D} \right)^3  \eqno(\ref{eq3-14}a)  $$
here $U$, $A$ and $S$ being each in unit of J. In the above, $D(\omega) =  V \omega^2 /2 \pi^2 c_1^3 = 3N\hbar \theta^2 / k_B \theta_D^3$ (with three longitudinal modes for each $K$ value) is the Debye density of states of $N$ oscillators; $\theta = {\hbar \w \ov k_B}$ and $\theta_D= {\hbar \w_D \ov k_B}={\hbar c_1 \ov k_B} ({6\pi^2N\ov V})^{{1\ov 3}}$; $<n(\w)>={1\over e^{\hbar \w /k_BT} -1}$ is the Planck distribution.  In the third expression of Eq. (\ref{eq3-12}a) and the third of  (\ref{eq3-13}a), $y= {\hbar \w \ov k_B T}$ and $y_D= {\hbar \w_D \ov k_B T}$; for $y_D \rightarrow \infty$: $\int^{y_D}_0 {y^3 \ov e^y -1} dy = \sum_{p=1}^{\infty} \int^{\infty}_0 y^3  e^{-py} dy = \sum_{p=1}^{\infty} {6\ov p^4}= {\pi^4 \ov 15}$. The momenton terms $\OUM$, $\OAM$ and $\OSM$ are omitted in the final expressions above.
\begin{eqnarray} \label{eq3-21}
&  \Us0 &= \Ph0  +   \int { 1 \ov 2} \hbar  \omega   \D(\omega) d\omega  
                 +  \int  { 1 \ov 2} \hbar \Omega  \D(\Omega) d\Omega  + \Delta_0.   \nnr  \\
&               &=  \Ph0 +  {3 \over 8} Nk_B (\theta_D + \gamma  \Theta_D) + \Delta_0, 
\end{eqnarray}
is the ground-state internal energy, where $\Ph0$ the ground-state potential energy; $<\e_n>_0={3\ov 8}k_B \theta_D$, $<\in_{{\cal N}}>_0 = {3\ov 8} k_B \Theta_D$ and $ \Delta_0$ are the zero-point kinetic energies per atom associated with the $\ub$, $\xib$, and $\Rb$ motions (Sec. \ref{Secz2}).  
With the $U_s$ above in (\ref{eq-CV0}) we get
\begin{eqnarray} \label{eq-CV} C_V \simeq     { Nk_B4 \pi^4 \over 5} \left({T \over \theta_D}\right)^3 ={Nk_B2 \pi^2 k_B^3 T^3 \over 15  (N/ V)  \hbar^3  c_1^3}, \qquad\qquad   (\ref{eq-CV0}a) \nonumber
\end{eqnarray}
with $C_V$ in unit of J/\oK. Finally, for the superfluid in a steady flow motion of velocity $\vb_s$, with the $S_s$ of (\ref{eq3-14}a) we obtain the entropy current
\begin{eqnarray} \label{eq-jS} \jb_S = \rho_s \vb_s S_s 
= Nk_B {4 \pi^4 \rho_s  \ov 15} \left( {T\ov \theta_D} \right)^3 \vb_s.   \end{eqnarray} 

With $\Us0=-7.2$\oK/atom and $c_1=239$m/sec as given previously, the above functions can be then parameterized. The resulting $U_s/Nk_B$, $A_s/Nk_B$, and $S_s/Nk_B$ correspond to the graphs in the $T \alt 0.6$ \oK \ region in FIG. \ref{fig-Tlmb} and these will be discussed in Sec. \ref{Secz10}. 
The resulting $C_V(T)$  (solid line in FIG. \ref{fig-Cv}a) shows a remarkable Debye $T^3$ behaviour; this has been used as a fed-back information in Sec. \ref{Secz2.1}. A least squares fit to the experimental data between  $0 \sim 0.6$ \oK \ using Eq. (\ref{eq3-31})$'$ gives $c_1 =241 $ m/sec and $\theta_D = 20 $ \oK. We furthermore observe that, the Debye region in the experimental $C_V$ ends at about 0.6\oK, which is in rough accordance with the usual $T^3$ law validity criterion,\cite{kittel} $T \le \theta_D / 50 \simeq 0.4 $ \oK. \ Anomalouslly, the $\lambda$ profile immediately above 0.6 (1) \oK \ results from the second order phase transition and should not be confused with a regular ending of the $T^3$ region. It also is anomalous that, at $T$ ($=$2.17 \oK) \ far below $\theta_D$ (20 \oK), \ the superfluid has already transformed to a normal liquid.

\subsection{The normal superfluid }  \label{Secz9.2}        \label{Secz9.C}
For the normal fluid of $N$ identical, distinguishable and weakly interacting classical He atoms, the Boltzmann statistical mechanics expressions can be readily written down. The total partition function is
\begin{eqnarray} \label{eq3-22}  
Z_n  =  e^{ -{U_{on} \over k_BT} }  {1 \over N!}  z^N,    \end{eqnarray}
where the $N$ factory reduces the identical distributions (due to the permutation among the $N$ particles) to be counted as one distribution. $U_{on}$ is similarly defined as $\Us0$ of Eq. (\ref{eq3-21}), and can in practice be taken as $\Un0=\Us0$, since at zero \oK \ it reduces to the same ground state as for the superfluid.
$z = \sum_{i_x,i_y,i_z=0}^{\infty} e^{- E_{i_x,i_y,i_z} /k_BT}$ 
and \mbox {$\epsilon_i = (i_x^2 + i_y^2 + i_z^2) h^2 / 8mV^{2/3}$} are the single atom partition function and kinetic energy. 
\begin{eqnarray} \label{eq-z}
 z = \int_0^{\infty} {\pi \over 4} \left({ 8m \over h^2}\right)^{3/2}
                  V \sqrt{E} e^{- E /k_BT} d E = {V \over \Lambda^3},
\end{eqnarray}
where $ \La = h / \sqrt{ 2\pi m k_B T}$ is the thermal de Broglie wavelength. 
Using  $z$ of (\ref{eq-z}) in Eq. (\ref{eq3-22}) gives $\ln Z_n = -\Us0/k_BT + N\ln z - N (\ln N - e ) = -\Us0/k_BT + N\ln ( Ve /N\Lambda^3)$, where $e=2.7183$.
Placing this in equations (\ref{eq3-12})-(\ref{eq3-15}) gives:  $U_n = \Us0 + {3 \over 2} N k_BT$ \ (Eq. \ref{eq3-12}b) and thus $E= {3 \over 2} N k_BT$, 
$ A_{n} = \Us0 + N k_B T \ln \left[ { N \over  Ve} { h^3 \over  (2\pi m k_B 
T)^{3/2}}  \right] $   \ (Eq. \ref{eq3-13}b), 
$S_n  =  - Nk_B \ln \left[ {\Lambda^3\ov e^{5/2}}  {N \ov V}\right] $ \ (Eq. \ref{eq3-14}b), 
and $C_{Vn}   = {3 \over 2}  Nk_B$ \ (Eq. \ref{eq3-15}b).

\subsection{The two fluid coexistent } \label{Secz9.3} \label{Secz9.D}
\subsubsection{ Two fluid fractions.} \label{Secz9.3.1}\label{Secz9.D1}  Consider now He II is heated slowly such that at each instant the liquid can be regarded as in (quasi) equilibrium; let this start from zero Kelvin where He II is a (pure) superfluid, and $V_s(0)=V$ and $N_s(0)=N$. At a given $T$, by Sec. \ref{Secz2.3} the fluid is a coexistent of the $s$ regions (Sec. \ref{Secz2}):  $V_s(T) = \sum_i V_{si}$, $N_s(T) = \sum_i N_i$, and the $n$ regions: $V_n(T) = \sum_i V_{ni}$, $N_n(T) = \sum_i N_{ni}$;  thus $V=V_s(T) + V_n(T)$ and $N=N_s(T) + N_n(T)$.  Furthermore, by Sec. \ref{Secz2.2} the $s$ to $n$ conversion involves the conversion of the superfluid bond $\Delta_b$ to the "van der Waals bond". The probability of the conversion is proportional to the Boltzmann factor, $ 1 /[ e^{\Delta_{sn} /k_BT} -1] \approx e^{-\Delta_{sn} /k_BT}$, $\Delta_{sn} $ being as given in Eq. (\ref{eq1-1}a). The density fractions of the two fluids of He II then follow to be
\refstepcounter{equation}
$$\displaylines{\label{eq3-28}  \hfill  x_n(T)   = { N_n \over N} = x_{on} e^{- {\Delta_{sn} \over k_B T }}   = e^{- {\Delta_{sn} \over k_B T }  (1 - T / T_o)  },         \hfill (\ref{eq3-28}a)   \cr  
  \hfill  x_s(T) = {N_s \ov N}= 1 - x_n = 1 -  e^{- {\Delta_{sn} \over k_B T } (1 - T / T_o)} \hfill (\ref{eq3-28}b)   
}$$
where $x_{n0} = e^{\Delta_{sn} \ov k_BT_o }$ is determined by satisfying the condition $x_n=1$ at $T=T_o$, $T_o$ being as defined in Sec. \ref{Secz4}. $x_s $ and $x_n$ here and the $f_s$ and $f_n$ of Sec. \ref{Secz4}, are related by: $x_s = f_s {\rho_s \ov \rho}$ and $x_n= f_n {\rho_n \ov \rho}$, and are basically equivalent since ${\rho_s\ov \rho}\approx {\rho_n\ov \rho} \approx 1$. In Eq. (\ref{eq1-1}a), $E(T) = {3\ov 2}k_BT$; and put $\dn \approx \dv$, then $\Delta_b + \dn \approx 8.6$ \oK, \ where $\Delta_b =-7.2 $ \oK /atom.\cite{simon:el:1950,london:1954}  
Eq. (\ref{eq1-1}a) is thus parameterized as:
$$ \Delta_{sn} =7.2 + 1.4 + 1.5 k_BT.              \eqno(\ref{eq1-1}a)^{\prime} $$
As to $T_o$, we here only look at its limits. By the free energy function of Sec. \ref{Secz10}, the onset of $n-s$ conversion would not occur until $A_s \le A_n$, which begins at $T \simeq 3.0$ \oK \ (see the context of Eq. \ref{eq3-39}). On the other hand, $T_o$ would not be lower than $T_{\la}$ which is 2.17 \oK \ from experiment. That is, $ 2.17 \le T_o \le 3.0$ \oK; \ the evaluation of $C_V$ in Sec. \ref{Secz9.D2} will show, optimally, $T_o=$ 2.6 \oK. \ With $T_o=2.17$, $2.6$ and 3.0 \oK \ in equations (\ref{eq3-28})-(\ref{eq3-28})$'$, and with the $\Delta_{sn}$ in (\ref{eq1-1})$'$, we obtain three sets of $x_n(T, T_o)$ and $x_s(T,T_o)$ functions (solid lines in FIG. \ref{fig-frac}b).  $T_o=2.6$ \oK \ yields indeed the optimum fractions (solid lines 3 and 3$'$ in the lower graph) which agree remarkably well with the $f_s(T)$ and $f_n(T)$ (dashed lines) obtained in Sec. \ref{Secz4}, and are bounded by the limit fractions with $T_o=2.17$ and 3.0 \oK\ (solid lines 1, 1$'$ and 2, 2$'$).
\input epsf  \begin{figure} [here] \begin{center} \leavevmode \hbox{%
\epsfxsize= 9 cm     \epsfbox{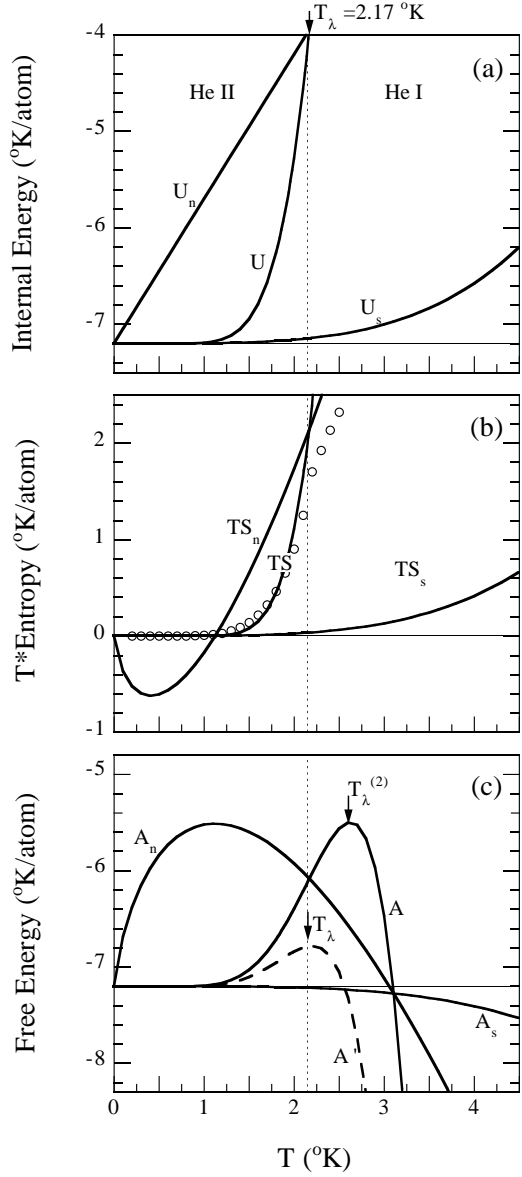}} \end{center} 
\caption{Theoretical thermodynamic functions of the superfluid ($s$), the normal fluid ($n$) and their coexistent He II (without subscript) (solid lines): (a) internal energies $U_s$, $U_n$, and $U$ as given  by Eqs. (\ref{eq3-12}a,b), (\ref{eq3-30})$''$,  (b) entropies $S_sT$, $S_nT$, and $ST$ as given by Eqs. (\ref{eq3-14}a,b) and (\ref{eq3-30xx})$''$,  and (c) free energies $A_s$, $A_n$, and $A$ as given by Eqs. (\ref{eq3-13}a,b) and (\ref{eq3-30x})$''$. Circles in (b) represent experimental calorimetric entropy data  \ref{wiebes:el,Kramers:1952,hill}.  
The maximum of $A$ of graph (c) yields the theoretical superfluid phase transition temperature $\Tl2$. $T_{\lambda} =2.17$ \oK \ is the experimental phase transition temperature at SVP. 
For the superfluid component of He II, the functions are each given by a canonical ensemble average over the states, with a Debye type of density of states, of the independent, indistinguishable single phonons, which are produced due to in real space the harmonic oscillations in the normal coordinates of the ($N$) independent, localized superfluid atoms relative to $\xib$. For the normal fluid, the functions are each a Boltzmann statistic mechanics result. The total functions result each as the weighted sum of the respective functions of the superfluid and the normal fluid, where the weighting function is given by the probability for the excitation of the superfluid bonds.  \label{fig-Tlmb} } \end{figure}

\noindent
\subsubsection{ Total thermodynamic functions.} \label{Secz9.3.2} \label{Secz9.D2} With the $x_s(T),x_n(T)$ being now known, the total thermodynamic functions of He II can be obtained as:
\begin{eqnarray} \label{eq3-30} && U  = x_s U_s + x_n U_n,  \\  
\label{eq3-30x} && A  = x_s A_s + x_n A_n, \\  
\label{eq3-30xx} && S  = x_s S_s + x_n S_n, \\ 
\label{eq3-31}  && C_{V}  = x_s C_{Vs} + x_n C_{Vn}  + {\pd x_s \ov \pd T} (U_s - U_n),   \quad {\rm and } \quad \\
 \label{eq3-21b}
 && {\bf j}_{Entr } = \rho_s  \vb_s S_s/Nk_B + \rho_n  \vb_n S_n/Nk_B.
\end{eqnarray}

\subparagraph{ Between 0 to 0.6 \oK. } \label{Secz9.3.3} \label{Secz9.D3} \  Here equations (\ref{eq3-28}) give $x_s(T) \simeq 1$ and $x_n(T) \simeq 0$, as we have already assumed so when representing the He II in this $T$ region with Eqs. (\ref{eq3-12}a)-(\ref{eq3-14}a) and (\ref{eq-jS}); compare FIG. \ref{fig-frac}b. 
Hence Eqs. (\ref{eq3-30})-(\ref{eq3-21b}) reduce to Eqs. (\ref{eq3-12}a)-(\ref{eq3-14}a) and (\ref{eq-jS}).
\subparagraph{  Between 0.6 ($\sim 1$) to 2.17 \oK \ (\Tlmb). } \label{Secz9.3.4} \label{Secz9.D4} \ Here a drastic conversion from the superfluid to the normal fluid occurs, with both of the fluids in Eqs. (\ref{eq3-30})-(\ref{eq3-21b}) having significant weights: $0 < x_s$, $ x_n$ $ < 1$. By Sec. \ref{Secz2.3} the two fluids of He II here aggregate in macroscopically large separate regions, thus the lattice statistical thermodynamics results (Sec. \ref{Secz9.B}) -- strictly each plus a higher order term $O_Y(T)$  ($Y U_s,A_s$, or $S_s$), and the Boltzmann statistics results (Sec. \ref{Secz9.C}) apply to each fluid regions, across the entire transition $T$ regime except in the immediate proximity of the critical point \Tlmb\ where the atomic dynamics of both the fluids become strongly long-range cooperative, for which our present treatment would not hold valid.  
 Using in Eqs. (\ref{eq3-30})-(\ref{eq3-31}) the thermodynamic functions of Eqs. (\ref{eq3-12}a)-(\ref{eq3-14}a) and (\ref{eq-jS}) for the superfluid terms, of (\ref{eq3-12}b)-(\ref{eq3-14}b) for the normal fluid terms, and the $x_s$ and $x_n$ of Eqs. (\ref{eq3-30})-(\ref{eq3-30})$'$, we obtain: 
$$ 
 U =  \Us0-  \left( \Us0-U_{on} \right)  e^{- {\Delta_{sn} \over k_BT}   (1- {T \over T_o})}  
+ Nk_B \left\{ {\pi^4 T^4 \over 5\theta_D^3} - 
      \left[  {\pi^4 T^4 \over 5\theta_D^3} - {3 \over 2}k_BT  \right]  e^{- {\Delta_{sn} \over k_BT} (1-{T \over T_o})} + O_U(T)\right\},
                                               \eqno(\ref{eq3-30})''   $$
$$ 
A  = \Us0 - (\Us0 - U_{on}) e^{-{\Delta_{sn} \over k_B T } (1 - {T \over T_o}) }  
+ Nk_B \left\{- {\pi^4 \over 15}    {T^4 \over \theta_D^3} + \left[  {\pi^4 \over 15} {T^4 \over \theta_D^3}  + T \ln \left({N \over V e} {h^3 \over (2\pi m k_BT)^{3/2}}\right)  \right] \right.  $$ $$ e^{ - {\Delta_{sn}  \over k_B T }  (1 - {T \over T_o}) }   \lf. + O_A(T) \right\},
                                                                      \eqno(\ref{eq3-30x})''   $$ $$
S =Nk_B  \left\{ {4 \pi^4 \over 15}   \left( {T\over \theta_B}\right)^3  
 - \left[ {4 \pi^4 \over 15}   \left( {T\over \theta_B}\right)^3  \right. 
 + \left. \ln \left(  { \Lambda^3 \over e^{5/2} } {N \over V} \right) \right] e^{- {\Delta_{sn}  \over k_B T } (1 - {T \over T_o}) } +O_S(T) \right\},
                                                                     \eqno(\ref{eq3-30xx})''  $$
$$  C_V = N k_B \left\{ {4\pi^4 \over 5} \left(  {T \over \theta_D }\right)^3  
      + \left[  { 3\over 2} \left(1+ {\Delta_{sn}\over k_BT}\right) -{4\pi^4  \over 5} \left({T \over \theta_D}\right) ^3  \left( 1- {\Delta_{sn} \over 4k_BT}\right) \right]              e^{- {\Delta_{sn} \over k_BT} ( 1 - {T \over T_{o}} )} \right\}.     \eqno(\ref{eq3-31})''    $$

We first evaluate $C_V$. Using in Eq. (\ref{eq3-31})$''$ the two limiting values 2.17 and 3 \oK \ for $T_0$ as discussed in the beginning of this section and $\Delta_{sn}$ of Eq. (\ref{eq1-1}b), we obtain two $C_V(T)$ limits (solid lines -1 and -2 in FIG. \ref{fig-Cv}b), which fall near to, actually enclosing in the middle, the experimental $\lambda$-profile (circles),  therefore defining a region of uncertainty in the theoretical $C_V(T)$. Within the limits, we find at $T_o=2.6$ \oK \ the $C_V(T)$ (solid line -3) optimally matches the experimental  $C_V$ data (circles). The remarkably good prediction of $C_V$ and similarly $x_s$ ($x_n$) earlier both with $T_o=$ 2.6 \oK, \ although being fortuitous in principle, can however find the following physical explanation. First, an effective $T_o$ intermediate between 2.17 and 3.1 \oK \ is expected to exist as discussed earlier. Second,  it can be shown that the $\la$-profile is primarily sensitive to the exponent, in which, the only other parameter $\Delta_{s}$, thus $\Delta_{sn}$, is given consistently by various experiments (cf. Sec. \ref{Secz2}).

The first order functions, Eqs (\ref{eq3-30})$''$-o(\ref{eq3-30xx})$''$ for the coexistent, Eqs. (\ref{eq3-12}a)-(\ref{eq3-14}a) for the pure superfluid and Eqs. (\ref{eq3-12}b)-(\ref{eq3-14}b) for normal fluid, can be parameterized, with the values for $\theta_D$, $\Ds$, and $T_0$ (2.6 \oK) \ as above and $U_{0s} = -7.2$ \oK/atom \ as used several times earlier. The resulting functions are graphically shown in FIG. \ref{fig-Tlmb}, a, b, c,  across the whole $T$ range of He II; 
for the entropy there exist experimental data (circles in FIG. \ref{fig-Tlmb}b),  \cite{wiebes:el,Kramers:1952,hill} with which the theoretical function $S(T)$ agrees satisfactorily, except for a noticeable suppression about the $T_{\Lambda}$ as can be expected.
We observe several as-expected features for $T<T_{\la}$: 1). The internal energy and hence the kinetic energy $<\e_n> = (U_s-\Us0)/Nk_B$, the entropy (at the high $T$ end) and the free energy of the pure superfluid are qualitatively lowered compared to those of the normal fluid, by $<\e_n> - E_{nr}= -(U_s - U_{n})/Nk_B \sim 3.1$,  $(S_s-S_n)/Nk_B \sim 2$, and $(A_s-A_n)/Nk_B =1$ \oK/atom \ respectively at $T$ just below 2.17 \oK. \ 2) From about $0.6$ \oK \ the total functions (of the coexistent) asymptotically approach to the corresponding functions of the pure superfluid with falling $T$, simultaneously He II approaches to a predominately pure superfluid: $x_s \rightarrow 1$.  3). The entropy of the normal fluid from 2.17 K reduces rapidly with falling $T$ from $T_{\la}$, and to a negative function in the regime below $\sim 0.6$ \oK, \ in which it becomes the significant source of the large positive $A_n$, or, of the normal fluid phase instability. Instead, the localized coherent oscillation scheme of the superfluid produces here a relatively high degree of disorder. 4). $U_s \simeq \Us0$ on the scale of $\Us0$.
5). With $\theta_D=20$ \oK \ obtained above, we have the zero-point  vibrational energy per atom of the $\ub$- motion (see the discussion below Eq. \ref{eq3-21}):
\begin{eqnarray} \label{eq-ena}
<\e_n>_0 /Nk_B = 7.5 \quad (^oK/{\rm atom}) 
\end{eqnarray}

\section{The superfluid phase transition and \Tlmb} \label{Secz10}  
Near $T_{\la}$, from FIG. \ref{fig-Tlmb} we observe that the first order functions of each pure fluid component all suffer discontinuous changes on transforming from one (normal) component to the other (superfluid).  However, as the phase transition involves a gradual conversion from one (the normal) component to the other (the superfluid), from $x_n(T_o)=1$  to $x_n(0)=0$, the first-order total thermodynamic functions of the coexistent, He II, are continuous; this is as expected for the second order phase transition here.

Futhermore, from FIG. \ref{fig-Tlmb} the onset of the $n$-$s$ conversion can clearly be seen  to occur near to 3.0 \oK, \ where a lower $A_n$ switches to a lower $A_s$:
\begin{eqnarray} \label{eq3-39}  A_s (\Tlm) - A_n (\Tlm) \le 0.  \end{eqnarray}
With the $A_s$ and $A_n$ given in (\ref{eq3-13}a) and (\ref{eq3-13}b),  (\ref{eq3-39}) becomes 
$$ e^{-19.2 \left(T/\theta_D\right)^3} \le  { nh^3 \ov e \sqrt{2 \pi m k_B T}^3}.               \eqno(\ref{eq3-39})'  $$
Since $19.2 (T/\theta_D)^3<< 1$ for the He II temperatures, we can write the left hand side of (\ref{eq3-39})$'$ as $e^{-19.2 (T/\theta_D)^3} \approx 1$ and re-ordering, we obtain  
\begin{eqnarray} \label{eq3-39b}
\xT1 \simeq  \left({N \ov V e}\right)^{2/3} {h^2 \ov 2 \pi m k_B}. \end{eqnarray}
It is noteworthy that this $\xT1$, as a direct statistical thermodynamic result for the two fluids of He II, is basically identical to the $T_{\la}$ given by London,\cite{london:1954,london:1938} although London's $T_{\la}$ was obtained, whilst lacking a direct thermodynamic scheme, by imposing a limitation in the thermal wavelength obtained for the normal fluid based on Boltzmann statistical mechanics. Placing now in Eq. (\ref{eq3-39b}) the experimental critical density of He II, $n={N\ov V} = 2.182 \times 10^{28}$ atoms/m$^3$, and the values for the other constants as previously specified, we obtain  
$$\Tlm^{(1)} \simeq 3.05  \quad  ^oK. \eqno( \ref{eq3-39b})'$$
\\
However, theoretically the initial $n$-$s$ conversion under the condition (\ref{eq3-39}), onsetting at $T^{(1)}_{\la}$, does not lead the total free energy $A$ to maximum until $T^{(2)}$, as can be seen in  FIG. \ref{fig-Tlmb}c (solid curve). It is the $A$ maximum which resembles the critical instability of the system in the second order phase transition, and where a qualitative $n-s$ conversion will occur. This critical condition formally writes:
\begin{eqnarray}  \label{eq3-40}   {\pd ^2 A \ov \pd T ^2} = 0,  \end{eqnarray} 
which expands as 
\begin{eqnarray}&& {\Delta_{sn} \over k_BT^2} ({\Delta_{sn} \over k_BT^2} - {2\over T}) (A_n-A_s) + {2\Delta_{sn} \over k_BT^2} {\partial \over \pd T}(A_n-A_s) \nonumber \\
 && \quad\quad + {\partial^2 \over \pd T^2} (A_n-A_s)=0.   \nonumber 
\quad\quad\quad\quad\quad\quad\quad\quad\quad (\ref{eq3-40})' \end{eqnarray}
Here the first order term of ($A_n-A_s$) represents just the condition (\ref{eq3-39}), and the first and second derivative terms of ($A_n-A_s$) are additional; (\ref{eq3-40})$'$  is a complex second order differential equation of $A_n-A_s$. Now instead of seeking an analytical solution of Eq. (\ref{eq3-40}), we can readily graphically locate, as we just actually did, the critical temperature from the $A$ maximum in FIG. \ref{fig-Tlmb}c to be: 
\begin{eqnarray} \label{eq3-40b} \Tl2 \simeq 2.6 \quad\quad ^oK.  \end{eqnarray}
Compared to $\xT1$, $\Tl2$ is seen to be lowered in value in closer agreement with the experimental value 2.17 \oK,  \ although with still a quite large discrepancy. The major error source may be the neglect  of a strong atomic correlation which anticipates to present already in the normal fluid in this $T$ range, in presumably a similar way as in its destination state -- the superfluid; indeed  previous studies\cite{wilks} have actually shown that the normal liquid He I deviates from ideal gas when nearing $T_{\la}$. Correlation will suppress the $A_n$ value, thereby yield a lowered value for $\Tl2$. If assuming across the transition $T$ region  $A_n$ is uniformly modified to $A_n'$ and hence $A$ to $A'$ and letting the maximum of $A'$ yield the experimentally observed $T_{\la}$, we find  $A_n'\approx A_n-0.7$ \oK \ and $A'=A-0.7$  \oK \ (dashed curve in FIG. \ref{fig-Tlmb}).

\begin{acknowledgments}
This work is supported by the Swedish Natural Science Research Council (NFR) and partly by the Wallenbergs Stiftelse (WS); JXZJ specifically acknowledges the fellowships from the NFR and WS. 
We acknowledge the valuable encouragement from Professor M. Springford,  Dr. P. Meeson,  Dr. J. A. Wilson,  Professor K. Sk\"old,  and Dr. O. Ericsson.  Since the outline of the (complete) theory of He II during June - October, 1998, we have had valuable opportunities of exchanging ideas with many of the international specialists in liquid helium, superconductivity and  condensed matter physics, many of whom have given valuable encouragement to this work; we give our deep acknowledgements to them here.  We have improved our model for the static structure of He II (initially proposed to retain an intermediate range ordering) due to the valuable criticism by A. Schofield (Cambridge), regarding the inability of a fluid in maintaining a long range ordering.
This paper has been finalized during the one year visit of JXZJ at the Link\"oping University. 
\end{acknowledgments}

\begin{appendix} \label{appn:J}
\section{  Distinction of conditions for "fluidity" and "diffusivity".}  

The preceding discussions  in effect imply that with the superfluid He II fluidity is not accompanied by any discernible atomic diffusion. In combination with the equations of motions and their solution in Secs. \ref{Secz5.2} and  \ref{Secz5.4}, and, based on the well-understood characteristics of the liquid and solid systems, we now elucidate that these two kinds of phenomena do not need co-occur. Fluidity is a phenomenon that any part of the fluid can move relative to another, in a flexible manner and under a small degree of shear stress. In respect of motion, the realization of a "flow" only requires the collective translation of all of the atoms concerned; there is nothing specific about whether the individual atoms diffuse in the same time. In respect of the atomic interaction, fluidity results from the absence of shear elasticity, or, the condition that the shear attraction between the atoms is so weak such that the system is unable to resist any shearing pressure. This character, in the matrix form of the interatomic force $\bf F$ below, is represented by that the shear (tangential) attraction force terms -- the off diagonal components -- are zero (cf. also Eq. (\ref{eq-force})): 
\begin{eqnarray} \label{eq:press:1a}
 {\bf F} = \left(
{\begin{array}{ccc}
F_{xx}{}^{r}+F_{xx}{}^{a}+ & 0          &         0 \\
     0    & F_{yy}{}^{r}+F_{yy}{}^{a} &         0 \\
     0    &      0     & F_{zz}{}^{r}+F_{zz}{}^{a}
\end{array}}
\right)          \nonumber \\
    \quad \quad ({\rm  \ liquid \ at \ rest});
\end{eqnarray}
or,
\begin{eqnarray} \label{eq:press:1b}
 {\bf F} =  \left(
{\begin{array}{ccc}
F_{xx}{}^{r}+F_{xx}{}^{a}  
& F_{xy}{}^{r} & F_{xz}{}^{r} \\
F_{yx}{}^{r} &  F_{yy}{}^{r} +F_{yy}{}^{a}
& F_{yz}{}^{r}\\
F_{zx}{}^{r} & F_{zy}{}^{r} & F_{zz}{}^{r}+F_{zz}{}^{a}
\end{array}}
\right),  \nonumber \\
({\rm  liquid \ in \ flow \ motion.})
\end{eqnarray}
 For the liquid at rest as of (\ref{eq:press:1a}), all the off-diagonal force terms are zero, and the diagonal components $F_{x} =F_{y} =F_{z}$ consist of the tensile attraction force  ($F_{\alpha \alpha}^{a}$) and repulsion force ($F_{\alpha \alpha}^{r}$),  where $\alpha =x,y,z$;  the equality of the diagonal terms represents the isotropy of the liquid.  When in flow motion as of  (\ref{eq:press:1b}), the shear repulsion force terms, i.e. the off-diagonals,  are finite and this leads to viscosity. On the other hand, as of Eq. (\ref{eq:press:1a}) or (\ref{eq:press:1b}), for the liquid either at rest or in flow motion, the tensile attraction force terms are finite and this represents that the atoms are attracted to each other (true for all liquids). The above  are to be contrasted to the force condition for an elastic solid:
\begin{eqnarray}\label{eq:press:1c}
{\bf F} =\left(
{\begin{array}{ccc}
F_{xx}{}^{r}+F_{xx}{}^{a}  \quad
& F_{xy}{}^{r} + F_{xy}{}^{a}   \quad
& F_{xz}{}^{r}+F_{xz}{}^{a} 
\\
F_{yx}{}^{r} +F_{yx}^{}{a}  \quad
& F_{yy}{}^{r}+F_{yy}{}^{a} \quad
& F_{yz}{}^{r} + F^{a}{}_{yz} 
\\
F_{zx}{}^{r}+F_{zx}{}^{a}   \quad
& F_{zy}{}^{r}+F_{zy}{}^{a}     \quad
& F_{zz}{}^{r} + F_{zz}{}^{a}   
\end{array}}   \right)         \nonumber \\
   \quad\quad ({\rm elastic \ body}). 
\end{eqnarray}
where, the diagonal terms are generally not zero, whence yielding the rigidity. Diffusion, on the other hand, is produced if the gradients of some of the tensile force terms -- the diagonal terms -- are finite; e.g. diffusion in $x$ direction results from $ \partial F_{xx} / \partial x \ne 0$ which leads to a density flux $D \partial F_{xx} / \partial x$;  cf. Eq. (\ref{eq8}.b). The tensile force terms exist for both solid and liquid, this feature corresponding to the characteristically finite values of the diagonals in all the above matrices. Hence diffusion can identically occur in both the systems; or, they may identically not occur if the kinetic energy of the atoms is low, roughly lower than the (tensile) attraction potential barrier between adjacent atoms.  It is true that, in regular fluids, a trivial shearing elasticity is often accompanied by a high mobility of the single atoms. This is nevertheless not so in the superfluid, where the many quantum atom correlation (Sec. \ref{Secz3}) produces a contrast in the tensile and shearing interactions, and the significantly large tensile attraction prevents the atoms from diffusion.
\end{appendix}


\begin{thebibliography}{150}
\bibitem{a} As simple cubic, or any other crystal structure, is not an exact representation of the actually disordered superfluid, hence the $a=3.3$ is only close but not equal to the lattice constant $a_{\rho}=3.6$ which one would obtain if assuming a simple cubic structure with the experimental density of He II, $ \rho =0.146$ g/cm$^3$ as input.
\bibitem{irreg} The fact that the experimentally observed well defined excitation energy $\w(K)$ makes a  compelling evidence that the distortion here cannot have been caused by the fluctuation in cage parameters as we unrealistically proposed earlier.
\end{thebibliography}
\end{document}